\documentclass{article}
\usepackage{tikz}
\usepackage{natbib}
\usepackage[T1]{fontenc}
\usepackage{geometry}
\usepackage{subfiles}
\usepackage{amsthm}
\usepackage{amsmath}
\usepackage{amssymb}
\usepackage{slashed}
\usetikzlibrary{cd}
\usepackage{bm}
\usepackage{physics}
\usepackage{mathtools}
\usepackage{subfiles}
\usepackage{color}
\usepackage{caption}
\usepackage{subcaption}
\usepackage{hyperref}
\usepackage{marginnote}
\usepackage{comment}
\usepackage{xcolor}
\usepackage{multirow}
\usepackage{colortbl}

\DeclarePairedDelimiter\floor{\lfloor}{\rfloor}

\tikzset{>=latex} 
\colorlet{myblue}{blue!65!black}
\colorlet{mydarkblue}{blue!50!black}
\colorlet{myred}{red!65!black}
\colorlet{mydarkred}{red!40!black}
\colorlet{veccol}{green!70!black}
\colorlet{vcol}{green!70!black}
\colorlet{xcol}{blue!85!black}
\tikzstyle{vector}=[->,very thick,xcol,line cap=round]
\tikzstyle{xline}=[myblue,very thick]

\usetikzlibrary{arrows.meta}
\usetikzlibrary{calc}

\DeclareMathOperator{\ad}{ad}
\DeclareMathOperator{\diag}{diag}

\newcommand{\ii}{\mathrm{i}}
\newcommand{\ve}{\varepsilon}
\newcommand {\zb} {{\bar z}}

\newcommand {\CalA} {\mathcal A}

\newcommand {\BR} {{\mathbb R}}
\newcommand {\BC} {{\mathbb C}}
\newcommand {\BP} {{\mathbb P}}
\newcommand {\BZ} {{\mathbb Z}}
\newcommand {\BT} {{\mathbb T}}
\newcommand {\CalC} {\mathcal C}

\newcommand {\CalD} {\mathcal D}
\newcommand {\CalE} {\mathcal E}

\newcommand {\CalG} {\mathcal G}
\newcommand {\CalH} {\mathcal H}

\newcommand {\CalM} {\mathcal M}

\newcommand {\CalO} {\mathcal O}

\newcommand {\CalR} {\mathcal R}

\newcommand {\CalV} {\mathcal V}

\numberwithin{equation}{section}
\newcommand\beq{\begin{equation}}
\newcommand\eeq{\end{equation}}

\title{\fontfamily{qcr}\selectfont\bf Yang-Mills flows for multilayered graphene}

\author{\fontfamily{put}\selectfont\it Vasilii Iugov, Nikita Nekrasov\\
\\
{\fontfamily{cmr}\selectfont Simons Center for Geometry and Physics}$^{\rm n}$,\\
and \\
{\fontfamily{cmr}\selectfont Yang Institute for Theoretical Physics}$^{\rm v,n}$, \\
\\
{\fontfamily{qcs}\selectfont Stony Brook University, Stony Brook NY 11794-3636}}

\begin{document}
\maketitle
\begin{abstract}

We clarify the origin of magic angles in twisted multilayered graphene using Yang-Mills flows in two dimensions. We relate the effective Hamiltonian describing the electrons in the multilayered graphene to the ${\bar\partial}_{A}$ operator on a two dimensional torus coupled to an $SU(N)$ gauge field. Despite the absence of a characteristic class such as $c_{1}$ relevant for the quantum Hall effect, we show that there are topological invariants associated with the zero modes occuring in a family of Hamiltonians. The flatbands in the spectrum of the effective Hamiltonian are associated with Yang-Mills connections, studied by M.~Atiyah and R.~Bott long time ago. The emergent $U(1)$ magnetic field with nonzero flux is presumably responsible for the observed Hall effect in the absence of (external) magnetic  field. We provide a numeric algorithm transforming the original single-particle Hamiltonian to the direct sum of ${\bar\partial}_{A}$ operators coupled to abelian gauge fields with non-zero $c_1$'s. Our gradient flow perspective gives a simple bound for magic angles: if the gauge field $A({\alpha})$ is such that the YM energy $\int_{{\BT}^{2}} {\rm tr} F_{A({\alpha})}^2$ is smaller than that of $U(1)$ magnetic flux embedded into $SU(2)$, then $\alpha$ is not magic. 

\end{abstract}

\section{Introduction}

Graphene is a two-dimensional carbon material, which can be built out of the ordinary graphite at room temperature. Its ``moving parts'' are well within the realm of non-relativistic quantum mechanics. Yet, a lot of interesting physics of graphene is surprisingly close, e.g. \cite{Schmitt:2022pkd},  to that of relativistic quantum field theory, as was anticipated forty years ago \cite{Semenoff:1984dq}. In this paper we shall investigate yet another aspect
of relativistic physics in $1+1$ dimensions
exhibited in graphene-made constructs. 

Specifically, we are interested in the so-called \emph{magic angles} occurring in the
studies of bi-layered graphene sheets \cite{CaoYuan2018}. Effective Hamiltonians describing the electron propagation in the background of two-layered graphene lattice were proposed in \cite{LopesDosSantos, Bistritzer:2011rxr}. These can be approximated by a two dimensional Dirac operator \cite{2011San-Jose}. Furthermore, neglecting the interlayer $AA$ coupling gives the chiral limit \cite{TarnopolskyKruchkovVishwanath}:
 \beq  
 {\CalH} = 
 \left( \begin{matrix} 0 & {\CalD} \\
 {\CalD}^{*} & 0 \end{matrix} \right)                           
\label{eq:effham}
\eeq
effectively described by the chiral Dirac operator
\beq  {\CalD} = {\bar\partial}_{\bar z} + A_{\bar z}  \label{eq:chiral}
\eeq
coupled to an $SU(2)$ gauge field 
\beq
{\bf A} = A_{z} dz + A_{\bar z} d{\bar z} \, , \ {\bf A}^{\dagger} = - {\bf A}\, , 
\eeq
with 
\beq
z = {\sf x} + {\ii}{\sf y}
\eeq
${\sf x,y}$ being coordinates on a plane with Euclidean metric
\beq
d{\sf x}^2 + d{\sf y}^2\ . 
\label{eq:flat_metric}
\eeq
More explicitly \cite{TarnopolskyKruchkovVishwanath}\footnote{One should note that the \eqref{eq:connection_formula} is not written in the global trivialization of the bundle $P$. One can easily pass to such a trivialization, at the expense of introducing a constant $\alpha$-independent diagonal term}, 
\beq
  A_{\bar z} = \frac{{\ii} {\alpha}}{2} 
  \left( \begin{matrix} 0 & U({\sf x,y} ) \\
U ( - {\sf x , - y} ) & 0 \end{matrix} \right) \label{eq:connection_formula} \eeq
is determined by a double-periodic potential $U({\sf x, y})$, whose structure is reviewed below, 
and the parameter $\alpha = \alpha ({\theta})$ a function of the twist angle. 

{}
The  non-abelian structure of \eqref{eq:effham}, \eqref{eq:connection_formula} is induced by inter-layer hopping amplitudes. 
It was observed that for some ``magic'' values of the twist angle $\theta$
the exactly flat bands are formed, i.e.  the Hamiltonian \eqref{eq:effham}
has a zero mode
\beq 
{\CalH}\, {\Psi}_{\bf k} = 0 \, , \ {\Psi}_{\bf k}(x+m, y+n) = e^{\ii \left( k_{x} m + k_{y} n \right)} 
{\Psi}_{\bf k}(x, y)\, , 
\label{eq:zeroeig}
\eeq
for any quasimomentum ${\bf k} = (k_x, k_y)$. 

There are proposals in the literature linking these flat bands to superconductivity \cite{2018Po, 2018Isobe}. 
Our goal is to provide a geometric picture explaining the flat band phenomenon, and a numerical algorithm for finding the zero modes
${\Psi}_{\bf k}$. 
As a by-product, we shall explain the observation of \cite{TarnopolskyKruchkovVishwanath} relating the zero modes of the effective
one-particle Hamiltonian to wavefunctions of an electron on the first Landau level, as well as the observations of \cite{Parhizkar:2024som} on the non-abelian nature of spatially inhomogeneous backgrounds with flatbands and perfect localization. Our explanation also sheds light on the observations of quantum anomalous Hall effect in twisted bilayered materials (including graphene) in the absence of background magnetic field Ref. \cite{PhysRevX.13.031037, ObsFQAHE, FQAHE, twist}. Likewise, the observation \cite{Popov:2020vhd} on the existence of the second solution of the Dirac equation, posessing a pole, follows trivially from the decomposition of the rank two vector bundle as a sum of two line bundles with opposite first Chern classes. 

We should stress that the graphene systems with similar behavior include monolayers with sublattice structures Ref. \cite{CanoDunbrack}. We shall collectively call these systems twisted multilayered graphene, TMG. 
We shall explain these observations using the language of a non-abelian gauge fields in two dimensions.  

A mathematical study of magic angles in the chiral model of TBG was done in \cite{Becker2021, Becker2022, becker2023degenerateflatbandstwisted}. These papers show the importance of symmetry for the appearance of magic angles. \cite{becker2023degenerateflatbandstwisted} also points out the existence of doubly-degenerate flat bands at certain values of $\alpha$. Our approach differs from the discussion in these papers in two key ways. First, we use the language of gauge theory and Yang-Mills flows that is new to the study of flat bands in TBG. It allows us to generalize the symmetry of TBG and develop an apparatus that could be useful in studying TBG-like systems. Second, when applied to TBG specifically, our method predicts flat bands of both multiplicity $q=1$ and $q=2$ at different points of complex parameter $\alpha$, and allows one to calculate the number of additional parameters needed to observe $q=4$ and higher. We are also able to show that magic angles of multiplicity $q=3n$ do not occur under any perturbation that preserves the symmetry. 

The main idea of our paper can be summarized in a couple of paragraphs. 

\vskip 1cm

For a given $\alpha$, the Eq. \eqref{eq:connection_formula} represents a point in the infinite-dimensional affine space $\mathcal A(P)$ of all connections (gauge fields) on a trivial $SU(2)$-bundle over the two dimensional torus $\mathbb T^2$. defined by taking the fundamental domain of the Moire structure and pairwise identifying its opposite edges, see the Fig. 1(a) below. This torus inherits the flat metric \eqref{eq:flat_metric}, which in turns endows it with a complex structure, where $z = {\sf x} + {\ii} {\sf y}$ is a local holomorphic coordinate. Globally, $z$ has two basic periods ${\omega}_{1}, {\omega}_{2}$, whose ratio ${\tau} = {\omega}_{2}/{\omega}_{1}$ determines the complex structure
of $\BT^2$, making it an elliptic curve ${\CalE}_{\tau}$. After a rescaling by $\omega_1$
the coordinate $z$ can be written as
\beq
z = x  + {\tau} y
\eeq
with $x,y$ defined up to the integer shifts (cf. Eq. \eqref{eq:hol_forms} below).

The affine space ${\mathcal A}(P)$ becomes a vector space once a trivialization $P \approx {\BT}^{2} \times SU(2)$ is fixed. ${\mathcal A}(P)$ is foliated by the orbits of the group ${\CalG}_{\BC}$ of complex gauge transformations:
\beq
A_{\bar z} \mapsto g^{-1} A_{\bar z} g + g^{-1} {\partial}_{\bar z} g
\label{eq:complex_gauge}
\eeq
where 
\beq
g = g(x,y) = \left( \begin{matrix} {\sf a}(x,y) & {\sf b} (x,y) \\ {\sf c}(x,y) & {\sf d}(x,y) \end{matrix} \right)\, , \ g^{-1}(x,y) = \left( \begin{matrix} {\sf d}(x,y) & -{\sf b} (x,y) \\ -{\sf c}(x,y) & {\sf a}(x,y) \end{matrix} \right)
\eeq
is the double periodic function of $(x,y)$ values in $SL(2, {\BC})$, i.e
${\sf a,b,c,d} \in {\BC}$, ${\sf ad - bc} = 1$. A classical problem in mathematical physics is to find the normal form of $A_{\bar z}$ with respect to \eqref{eq:complex_gauge}. 

Some connections in $\mathcal A(P)$ admit a non-singular (normalizable on $\mathbb T^2$) section $\psi$ annihilated by the chiral Dirac operator \eqref{eq:chiral}
\beq
{\CalD} {\psi} = 0 
\label{eq:dzm}
\eeq
Mathematically, the Eq. \eqref{eq:dzm} is the connection between the differential and algebraic geometric aspects of Yang-Mills theory in two dimensions \cite{AtiyahBott0}. Namely, the complex structure on ${\BT}^{2}$ (which makes it an elliptic curve ${\sf E}_{\tau}$) and connection on $P$ defines the structre of a holomorphic rank two vector bundle ${\CalE}$ on the associated  
vector bundle $E$. The solution $\psi$ to Dirac equation is the global holomorphic section of ${\CalE}$: 
\beq
{\psi} \in H^{0}({\sf E}_{\tau}, {\CalE})
\eeq
defining a linear subbundle $L = {\BC}{\psi} \subset E$.

The solvability of \eqref{eq:dzm} is preserved by the action of the transformations \eqref{eq:complex_gauge}.

Most choices of $\bf A$ are unstable in the sense that the zero mode exists only for some isolated value of the quasimomentum $\bf k$. Curiously, algebraic geometers call the associated vector bundles stable.  However, there are rare connections in ${\mathcal A}(P)$ for which \eqref{eq:dzm} admits zero modes for any quasimomentum $\bf k$. The so-called flatbands appear when a solution of \eqref{eq:dzm} has $q>0$ zeroes. In this case, the quasimomentum can be changed by multiplying $\psi$ by a meromorphic function. Let $z_1, \dots,z_q\in \mathbb T^2$ be the zeroes of $\psi$. As an exercise, one can check that up to a constant factor there is a single periodic meromorphic function on the torus $f_{z_1,\dots,z_q}^{\tilde z_1,\dots,\tilde z_q}(z)$ with poles $z_1, \dots, z_q$ and zeroes $\tilde z_1, \dots, \tilde z_q$ as long as $z_1+\dots+z_i=\tilde z_1+\dots +\tilde z_i$. This function can be constructed explicitly as
\beq
f_{z_1,\dots,z_q}^{\tilde z_1,\dots,\tilde z_q}(z) = \prod_{i=1}^q \frac{\theta_1(z-\tilde z_i;\tau)}{\theta_1(z-z_i;\tau)}
\eeq
There is a $q$-dimensional space of solutions of \eqref{eq:dzm} (we give more details about this in the section 3.2 where we study the similar problem for $q \to 2q$). Moreover, through shifting one of the zeroes as
\beq
\psi_{\mathbf k} (z, \bar z)= \frac{\theta_1 \left(z-z_1-\frac{1}{2\pi \ii}\frac{\sqrt{3}}{2}\left( \delta k_x + {\ii}\delta k_{y} \right); \tau \right)}{\theta_1(z-z_1; \tau)}e^{{\ii}\delta k_x z} \psi_0(z, \bar z)
\eeq 
we can change the quasimomemtum $\bf k\mapsto \bf k+\delta \bf k$. Thus, $q$ linearly independent solutions exist for every point in the Brillouin zone. The integer $q=\dim \mathcal D$ can be associated to every orbit of the complexified gauge group $\mathcal G_\mathbb C$  acting in the space $\mathcal A(P)$. Thus, the study of magic angles reduces to the problem of studying the orbits of $\mathcal G_\mathbb C$ in $\mathcal A(P)$. Varying $\alpha$ as a complex parameter defines a two dimensional surface in $\mathcal A(P)$. We are therefore interested in knowing whether this surface intersects the locus of connections with flatband(s).

Our idea of how to determine whether ${\CalD}$ in \eqref{eq:effham} for a given $A \in {\mathcal A}(P)$
admits such a zero-mode(s) is to employ the Yang-Mills flows, the gradient flows of the Yang-Mills action functional $S_{YM}$ on $\mathcal A(P)$. The critical subspaces ${\CalC}_q$ of $S_{YM}$ are labelled by an integer $q\ge 0$. They correspond to the reduction of the structure group of the initial $SU(2)$-bundle to its maximal torus $U(1)$. The $U(1)$-bundles over $\mathbb T^2$ are classified by the first Chern class, the flux $q$. The associated rank two vector bundle $E$ with the fiber ${\BC}^{2}$
decomposes as a direct sum of two non-trivial line bundles with the first Chern classes $q$ and $-q$. Connections that flow to ${\CalC}_q$  form a cell ${\mathcal A}_q \supset {\CalC}_q$ in the Atiyah-Bott decomposition. They satisfy $\dim_{\mathbb C} \ker {\mathcal D}_{A}=q$, thus admitting a flat band of degeneracy $q$. Two-dimensional gauge theory and YM flows are reviewed in section $\bf 2$.

We should stress that in gauge theory usually one encounters the spaces
${\CalM}_{q} = {\CalC}_{q}/{\CalG}$, the moduli spaces of Yang-Mills connections. These are finite dimensional (orbi)folds.

Given that the critical loci $\mathcal C_q$ of the YM flow are essential for understanding magic angles, we proceed to study them and their neighborhoods in more detail in section $\bf 3$. In particular, we show that the subspace $T_{\bf A_0}^{+} {\mathcal A}(P)$ of relevant perturbations (which we define) in the tangent space $T_{\bf A_0} {\mathcal A}(P)$ to a critical connection ${\bf A}_0 \in \mathcal C_q$ is isomorphic to $\mathbb C^{2q}$. Therefore, the complex codimension of ${\mathcal A}_q$ is equal to $2q$. Were $\mathcal A_{\bar z}(\alpha)$ of the Eq. \eqref{eq:connection_formula} a generic one-complex-parametric family of connections, it would not intersect a $2q$-codimensional space even for $q=1$.

Luckily, there is a resolution to this apparent mismatch. All connections in the family \eqref{eq:connection_formula} are invariant under some action of a finite group $\Gamma$,
covering the action by rotations of the base torus ${\BT}^{2}$. 
In section $\bf 4$ we study the spaces of  connections on the trivial $SU(2)$-bundle that have discrete  symmetries of that kind. With respect to $\Gamma$-action, the subspace of relevant perturbations at a $\Gamma$-invariant critical point ${\bf A}_0 \in {\mathcal C}^{\Gamma}_q$ decomposes into subspaces with different charges under said symmetry. The codimension of the cell ${\mathcal A}^{\Gamma}_q \subset {\mathcal A}(P)^{\Gamma}$, being equal to the dimension of the space of $\Gamma$-invariant relevant deformations, is thereby smaller. This helps us to explain the phenomenon of magic angles in the next section.

In section $\bf 5$, we focus our attention to the family  \eqref{eq:connection_formula} of gauge fields appearing in TBG. Using direct numerical computation, we find that all of the "magic angles" in the complex $\alpha$ plane correspond to points in ${\mathcal A}_1$. Moreover, we use the math developed in section $\bf 4$ to show that the appearance of magic angles is stable under perturbation of the connections that preserve $\Gamma$. 

In section $\bf 6$ we present our conclusions.

{\bf Acknowledgements.} We thank G.~Tarnopolsky and F.~Popov for discussions. We also  thank IHES for hospitality in 2021-2024, while part of this work has been done. NN thanks ICTP (Trieste), Institut Mittag-Leffler (Stockholm) and Uppsala University for their hospitality. The results of this work were first presented at Strings-Math'2024.  Research is partly supported by NSF PHY Award 2310279.

\section{Yang-Mills flow}

In this section we recall the definition of Yang-Mills flow, a gauge-theoretic analogue of the celebrated Ricci flow. Just as the (modified) Ricci flow describes an approximation to the renormalization group flow for a two dimensional sigma model \cite{Friedan:1980jm}, the Yang-Mills flow is an approximation to boundary renormalization group flow. 

Mathematically it has been defined and studied long time ago in two dimensions \cite{AtiyahBott, Donaldson}, and, in some special cases, more recently in higher dimensions \cite{Jacob}

\subsection{General notions}

Consider a principal $G=SU(N)$-bundle $P$ over a Riemannian manifold $M$. Let ${\mathcal A}_{P}$ be the space of $G$-connections on $P$. For each connection ${\bf A}\in \mathcal A(P)$, there is an associated curvature $F_{\bf A} \in \Omega^2 (M, \ad P)$. In local trivialization we can express $F_{\bf A}$ as
\begin{equation}
    F_{\bf A} = \dd {\bf A} + {\bf A}\wedge {\bf A}
\end{equation}
Define the Yang-Mills functional to be
\begin{equation}
    S_{\rm YM}({\bf A}) = - \int_M \Tr( F_{\bf A} \wedge \star F_{\bf A}) 
\end{equation}

The infinite dimensional manifold  $\mathcal A(P)$ of gauge fields has Euclidean $L^2$-metric. On the tangent space $T_{\bf A}\mathcal A(P)$ it is given by
\beq
\label{eq:metric}
    (\delta A_1, \delta A_2) = -\int_M \Tr(\delta A_1 \wedge \star \delta A_2)
\eeq
Given the functional $S_{\rm YM}$ and the metric \eqref{eq:metric}, we can define the gradient vector field
${\nabla} S_{\rm YM}$ on $\CalA (P)$. The Yang-Mills flow is the solution to the associated steepest descent equation: 
\begin{equation}
\label{eq:YMflow}
    \dot {\bf A} = \dd_{\bf A}^{*} F_{\bf A} + \dd_{\bf A} \chi
\end{equation}
where $\dd_{\bf A}$ is the covariant derivative in $\ad P$ associated with the connection ${\bf A}$, and ${\chi} = \chi({\bf A})$ is an arbitrary function $\mathcal A(P)\mapsto \ad P$. The term $\dd_{\bf A} \chi$ reflects the fact that a local trivialization of a connection is only defined up to a gauge transformation\footnote{We stress here that the $SU(2)$ gauge field the electrons are interacting with is not the dynamical gauge field, such as the one mediating weak force, for example. Although the gauge transformations of $\bf A$ do not change the nature of flatbands, the fact the system is composed out of two physical sheets of graphene gives a preferred trivialization of $P$.}  We choose later $\chi = 0$, thus ensuring that ${\dot{\bf A}}$ is orthogonal to the orbit of gauge group passing through $\bf A$, i.e. 
\beq
\dd_{\bf A}^{*} {\dot{\bf A}} = 0
\eeq
When ${\rm dim}(M) =2$ the metric \eqref{eq:metric} depends only on the conformal structure on $M$. 
Using the fact that $\dd_{\bf A}^{*} = \star \dd_{\bf A} \star$, and that $\star$ acts as complex structure on $1$-forms, $\star {\Omega}^{1,0}(M) = {\ii} {\Omega}^{1,0}(M)$, $\star {\Omega}^{0,1}(M) = - {\ii} {\Omega}^{0,1}(M)$, we conclude:
\beq
{\dot {\CalA}_{\bar z}} = D_{\bar z} \left( {\chi} - {\ii} {\phi} \right) \, , \ 
\label{eq:01gauge}
\eeq
where
\beq
{\phi} = \star F_{\bf A}
\label{eq:curv}
\eeq
Thus the flow \eqref{eq:YMflow} proceeds by complex gauge transformations:
\beq
{\CalA}_{\bar z}(t) = g(t)^{-1} {\CalA}_{\bar z}(0) g(t) + g(t)^{-1} {\bar\partial}_{\bar z} g(t)
\, , \eeq
where $g(t)$ solves
\beq
{\dot g} \, = \, g \left( {\chi} - {\ii} {\phi} \right) \ . 
\eeq
The flow \eqref{eq:YMflow} stops, $\dot{\bf A} = 0$, at the critical points of $S_{\rm YM}$. There are two classes of such points: the ones, corresponding to $F_{\bf A} = 0$ (flat connections), and the ones where
$F_{\bf A} \neq 0$ yet the Yang-Mills equations 
\beq
{\dd}_{\bf A}^{*} F_{\bf A} = 0 
\label{eq:yme}
\eeq
are satisfied. These are usually called the Yang-Mills connections. When the base manifold is a two-sphere, and the gauge group $G = SU(N)$ with $N \to \infty$ the combinatorics of YM connections plays an important role in the phase structure of planar theory \cite{Gross:1994ub}.
Let us study the structure of the flow \eqref{eq:YMflow} in the neighborhood of a critical point.

In what follows we shall mostly work over a two-torus $M = {\BT}^{2}$. The above mentioned complex structure on the space of $1$-forms (and, correspondingly, gauge fields) is associated with a choice of complex structure on $M$. The two-torus endowed with the complex structure is called the elliptic curve, and will be denoted by
$\CalE_{\tau}$. The eigen $1$-forms of $\star$ with the eigenvalue $\pm \ii$ are
\beq
dz = dx + {\tau} dy\, , \ d{\zb} = dx + {\bar\tau} dy\, , 
\label{eq:hol_forms}
\eeq
respectively. The parameter ${\tau}$, ${\rm Im}{\tau} > 0$ determines the complex structure of ${\CalE}_{\tau}$. The group $PSL_{2}({\BZ})$
acts on $\tau$ via
\beq
{\tau} \mapsto \frac{a {\tau} +b}{c{\tau}+d}
\eeq
This action doesn't change the complex structure of ${\CalE}_{\tau}$, only a choice of the basis
of flat coordinates $(x,y) \mapsto (d x + b y, a y + c x)$.  
The coordinates $(x,y)$ are related to the orthonormal coordinates ${\sf x,y}$ via:
\beq
{\sf x} = x + {\tau}_{1} y\, , \ {\sf y} = {\tau}_{2} y
\eeq
\subsection{Critical points in the two dimensional case}

In this section we classify the critical points of $S_{\rm YM}$. Let us work in a gauge in which $\phi$ \eqref{eq:curv} is diagonal. Let 
\beq
\phi = \diag( \phi_1, \dots, \phi_n) \ , 
\eeq 
where $\sum_i \phi_i = 0$. In that gauge, the Yang-Mills equations reduce to
\begin{equation}
\begin{split}
    (\phi_i - \phi_j) {\bf A}_{ij} = 0 \, , \ i \neq j\\
    \dd \phi_i = 0
\end{split}
\end{equation}
Thus, the curvature eigenvalues $\phi_i$ are constant and the connection $\bf A$ is block-diagonal, associated with a partition
\beq
N = d_{1} + \ldots + d_{l}
\eeq
of $N$ into $l$ parts, corresponding to the sets of equal eigenvalues of $\star F$:
\begin{equation}
    {\bf A} = \bigoplus_{k=1}^{l} {\bf A}_{k} 
\end{equation}
Each block $\bf A_k$ is a sum of the scalar component $a_k$, a $U(1)$ connection of constant curvature $\phi_k$ and a flat $SU(d_k)$ connection $\CalA_k$:
\beq
{\bf A}_{k} =  {\bf 1}_{d_{k}} \cdot a_{k} + {\CalA}_{k} 
\eeq
On a compact Riemann surface $\Sigma$, the curvature $da_k$ is an integral two-form:
\beq
\int_{\Sigma} da_k = 2\pi\ii q_k\, .\ 
\eeq
The fluxes $q_1, \ldots, q_l \in {\BZ}$ are constrained by:
\beq
\sum_{k=1}^{l} d_{k} q_{k} = 0 \  . 
\eeq
For the $SU(2)$ gauge group, the critical manifolds $\mathcal C_{q}$ are thus labeled by a single integer $q\ge 0$. Connections with $q=0$ have zero curvature and form the space of flat connections $\mathcal C_{0}$. The quotient ${\mathcal M}_{0} = {\CalC}_{0}/{\CalG}$ by the action of the gauge group $Maps({\BT}^{2}, SU(2))$ is a finite dimensional orbifold
\beq
{\CalM}_{0} \approx {\BT}^{2}/{\BZ}_{2}
\eeq
parametrized by the eigenvalues $e^{\ii \alpha}, e^{\ii\beta}$ of the commuting monodromies of the connection around the $1$-cycles of the torus, up to their simultaneous inversion $(e^{\ii \alpha}, e^{\ii\beta}) \sim (e^{-\ii \alpha}, e^{-\ii\beta})$. As a complex manifold 
\beq
{\CalM}_{0} \approx {\CalE}_{\tau}/{\BZ}_{2} \approx {\BC\BP}^{1}
\eeq
with the ${\BC\BP}^1$ parameterized by the Weierstrass function ${\wp}(z) = {\wp}(-z)$, $z = \frac{1}{2\pi} ( {\alpha} + {\tau}{\beta})$.

For $q > 0$ we can choose a gauge where
\begin{equation}
\label{eq:u1_constant_curvature}
    {\phi} = \frac{2 \pi i q}{S_{\BT}} \sigma_3, \quad \mathbf A = i a \sigma_3
\end{equation}
with $\dd a = \frac{2 \pi  q}{S_{\mathbb T}}$ and $S_{\mathbb T}$ the area of the torus.
Up to remaining gauge transformation, in coordinates $(x, y) \in [0, 1]^2$
\begin{equation}
\label{eq:linear_gauge}
    a(x, y) = -2 \pi q y \dd x +  \left( b_x \dd x + b_y \dd y \right)
\end{equation}
with $(b_x, b_y) \in ({\BR}/2\pi{\BZ})^2$, as the shifts $b_{x} \mapsto b_{x}+2\pi m_x$, $b_{y}\mapsto b_{y}+2\pi m_y$, with integer $m_x, m_y$ are actually gauge transforms. Thus, the quotient ${\CalM}_{q}$ of the critical manifold ${\CalC}_{q}$ by the gauge group is a two-torus 
\beq
{\CalM}_{q} \approx \mathbb T^2 \, , \ q > 0
\eeq
 parametrized by $b_x, b_y$ mod $2\pi$.  As in the case $q=0$, 
 the exponents $e^{\ii b_x}, e^{\ii b_y}$ are the eigenvalues of the monodromies of the critical connection $\bf A_0$ around certain loops
 on ${\BT}^2$. However, unlike the $q=0$ these monodromies change when the loops are continuously deformed. 
 
 \subsection{$S_{\rm YM}$ near a critical point}

Let ${\bf A}_0$ be a critical point of the YM functional, $\dd_{{\bf A}_0} \star F_{{\bf A}_0}=0$. Let 
\beq
{\bf A} = {\bf A}_0 + {\ve} {\delta \bf A}
\label{eq:close}
\eeq
be a nearby gauge field, ${\ve} \to 0$. 

We can expand:
\beq
S_{\rm YM}({\bf A}) = S_{\rm YM}({\bf A}_{0}) + {\ve}^{2} \int_{{\BT}^{2}} {\rm tr}\, {\delta}{\bf A} \wedge \star \left( {\mathfrak{D}}_{{\bf A}_0}({\delta \mathbf A}) \right) + \ldots
\label{eq:sym2}
\eeq
where the linear operator (the Hessian of $S_{\rm YM}$) ${\mathfrak{D}}_{{\bf A}_0} \in End(T_{{\bf A}_0} {\CalA}(P))$ is defined by:
\beq
{\mathfrak{D}}_{{\bf A}_0}({\delta \mathbf A}): = \star \dd_{{\bf A}_0} \star \dd_{{\bf A}_0} {\delta \mathbf A} - [ \star F_{{\bf A}_0}, \star {\delta \bf A}] 
\label{eq:hess}
\eeq
The gauge group action on $\CalA (P)$ translates to
\begin{equation}
\label{eq:gauge_transformation_near_critical_point}
\begin{split}
    {\bf A}_0 &\mapsto g^{-1} {\bf A}_0 g + g^{-1} \dd g \\
    {\delta \bf A} &\mapsto g^{-1} {\delta \bf A} g
\end{split}
\end{equation}
Infinitesimally, the gauge transformations $g(x) = 1 + {\ve}\, {\xi}(x) + \ldots$, with ${\xi}(x) \in Lie G$, act on ${\delta}{\bf A}$, 
defining a subspace $T^{\rm gauge}_{{\bf A}_{0}} {\CalA}(P) \subset
T_{{\bf A}_{0}} {\CalA}(P)$ of the tangent space -- the tangent space to the gauge orbit pasing through ${\bf A}_{0}$: 
\beq
T^{\rm gauge}_{{\bf A}_{0}} {\CalA}(P) = \{ {\delta{\bf A}} = {\dd}_{{\bf A}_{0}} {\xi} \, \} \approx {\rm Maps}({\BT}^{2}, LieG)/H^{0}_{{\bf A}_{0}}
\label{eq:tangorbit}
\eeq
where $H^{0}_{{\bf A}_{0}}$ is the tangent space to the group $Aut({\bf A}_{0})$ of automorphisms of ${\bf A}_{0}$ -- 
the group of gauge transformations that leave ${\bf A}_0$ invariant, i.e.  $g = g(x)$, such that $\dd_{{\bf A}_0} g = 0$.

\subsection{The relevant directions of flow near a critical point}

Substituting \eqref{eq:close} into the steepest descent equation \eqref{eq:YMflow}, we obtain the following linear equation on ${\delta \bf A}\in \ad P$:
\begin{equation}
    \frac{d}{dt} {\delta \mathbf A} = {\mathfrak{D}}_{{\bf A}_0}({\delta \mathbf A})
    \label{eq:YM_flow_near_critical_point}
\end{equation}
as ${\ve} \to 0$. 

We call  the \emph{relevant, marginal, irrelevant directions} in $T_{{\bf A}_0} {\mathcal A}(P)$, respectively, the eigenspaces $T_{{\bf A}_0}^{+, 0, -} {\mathcal A}(P)$ of ${\mathfrak{D}}_{{\bf A}_0}$ with positive, zero, negative eigenvalues of $\mathfrak{D}_{{\bf A}_0}$, respectively.
Since ${\mathfrak D}_{\bf A_0}$ is a symmetric operator, the tangent space $T_{\bf A_0}{\CalA}(P)$ splits as an orthogonal direct sum:
\beq
T_{\bf A_0}{\CalA}(P) = T^{+}_{\bf A_0}{\CalA}(P) \oplus T^{0}_{\bf A_0}{\CalA}(P) \oplus T^{-}_{\bf A_0}{\CalA}(P)
\eeq

Let ${\delta \bf A}$ be an eigenfunction of $\mathfrak{D}_{{\bf A}_0}$,
\begin{equation}
\label{eq:YM_flow_near_critical_point}
    \lambda {\delta \bf A} = {\mathfrak{D}}_{{\bf A}_0} {\delta \bf A} = \star \dd_{{\bf A}_0} \star \dd_{{\bf A}_0} {\delta \bf A} - [\star F_{{\bf A}_0}, \star {\delta \bf A}]
\end{equation}
Applying $\dd_{{\bf A}_0} \star$ to \eqref{eq:YM_flow_near_critical_point}, one obtains
\begin{equation}
    \lambda \dd_{{\bf A}_0} \star {\delta \bf A}  = - [F_{{\bf A}_0}, \star \dd_{{\bf A}_0} {\delta \bf A} ] - [\dd_{{\bf A}_0} {\delta \bf A} , \star F_{{\bf A}_0}] = 0
\end{equation}
Thus, for relevant or irrelevant ${\delta} {\bf A}$, $\dd_{{\bf A}_0} \star {\delta \bf A}  = 0$, in other words $T_{{\bf A}_0}^{+, -} {\mathcal A}(P)$ are orthogonal, w.r.t the metric, to the gauge orbit passing through ${\bf A}_0$. 

Obviously, the tangent space $T^{\rm gauge}_{{\bf A}_{0}} {\CalA}(P)$ to the gauge orbit passing through ${\bf A}_{0}$ belongs to $T_{{\bf A}_0}^{0} {\mathcal A}(P)$: 
\beq
T^{\rm gauge}_{{\bf A}_{0}} {\CalA}(P) \subset T^{0}_{{\bf A}_{0}} {\CalA}(P)
\eeq
Now, let ${\delta}{\bf A}$ be an eigenvector of ${\mathfrak D}_{\bf A_0}$ with eigenvalue $\lambda$. Applying $\dd_{{\bf A}_0}$ to \eqref{eq:YM_flow_near_critical_point}, one obtains
\begin{equation}
    \dd_{{\bf A}_0} \star \dd_{{\bf A}_0} \star \dd_{{\bf A}_0} \, {\delta \bf A} = \lambda \dd_{{\bf A}_0} \, {\delta \bf A}
\end{equation}
Therefore, unless $\dd_{{\bf A}_0} \, {\delta \bf A} = 0$, $\lambda$ is also an eigenvalue of $ \dd_{{\bf A}_0} \star \dd_{{\bf A}_0} \star$ acting on $\ad P$. The operator $ \dd_{{\bf A}_0} \star \dd_{{\bf A}_0} \star$ is negatively defined; thus, for $\lambda > 0$ we necessarily have $\dd_{{\bf A}_0} {\delta \bf A} = 0$. Thus, the first term in \eqref{eq:YM_flow_near_critical_point} cancels  and we are left with
\begin{equation}
\label{eq:commutator_eigenvalue}
    \begin{split}
        \lambda^2 \, {\delta \bf A} \, =\,  \lambda [\star {\delta \bf A}, \star F_{{\bf A}_0}] = - [[{\delta \bf A}, \star F_{{\bf A}_0}], \star F_{{\bf A}_0}]
    \end{split}
\end{equation}
It is easy to solve \eqref{eq:commutator_eigenvalue} in a gauge where $\star F_{{\bf A}_0}$ is diagonal.  An eigenvector ${\delta \bf A}$ with $\lambda \neq 0$ is given by an off-diagonal matrix valued one-form.

\section{$SU(2)$ case on a torus}

\label{sec:torus_YMflow_theory}

\subsection{Near a critical point}

Consider the neighborhood of a point on a critical manifold of $SU(2)$ connections with $q>0$. In the diagonal gauge, following the above discussion adapted to the $N=2$ case we solve \eqref{eq:commutator_eigenvalue} with
\begin{equation}
\label{eq:solutions_in_diagonal_gauge}
    {\delta \bf A} = \begin{pmatrix}
    0 & \xi \dd z \\
    -\xi^\star \dd \bar z & 0 \end{pmatrix}\quad \text{or} \quad {\delta \bf A}  = \begin{pmatrix}
    0 & \xi^\star  \dd \bar z \\
    -\xi \dd  z & 0 \end{pmatrix}
\end{equation}

One of the solutions \eqref{eq:solutions_in_diagonal_gauge} corresponds to $\lambda = 2 f$, and the other one to $\lambda = -2f$. Equation $\dd_0 {\delta \bf A}=0$ reduces to 
\begin{equation}
\label{eq:xi_equation}
    (\bar \partial + 2a_{\bar z}) \xi = 0
\end{equation}
where
$a_{\bar z}d{\bar z} + a_{\bar z}^{*}dz = a(x,y)$ given in \eqref{eq:linear_gauge}. For a connection of curvature \eqref{eq:u1_constant_curvature} there are exactly $2q$ solutions to that equation (reviewed in the next section), giving $2q$ relevant directions near a critical point. 

In the $SU(2)$ case the gauge transformations \eqref{eq:gauge_transformation_near_critical_point} that leave ${\bf A}_0$ invariant are the constant diagonal matrices $g = e^{\ii u {\sigma}_{3}}$. They act on connections of the form \eqref{eq:solutions_in_diagonal_gauge} by phase rotating $\xi \mapsto e^{2\ii u} {\xi}$. In addition, rescaling $\xi \mapsto e^{v} {\xi}$ simply changes the parametrization of the gradient flow. This means that the solutions of \eqref{eq:xi_equation} are only important up to a constant complex factor. Thus, the relevant physical deformations of a critical point of flux $q$ are parametrized by $\mathbb{CP}^{2q-1}$.

\subsection{Explicit formulas for relevant operators}

In the gauge 
\begin{equation}
\label{eq:convenient_gauge_arbitrary_tau}
    a_q = -\frac{\pi}{2\Im \tau} (qz \dd \bar z - q\bar z \dd z+ b \dd \bar z - \bar b \dd z) = a_{z} dz + a_{\bar z} d{\bar z} 
\end{equation}
the solution to  \eqref{eq:xi_equation} can be explicitly presented as 
\begin{equation}
    \xi_q(z, \bar z) = \exp(-\frac{\pi q}{\Im \tau} z(\bar z - z) + \frac{\pi}{\Im \tau}(-b\bar z + b z)) f(z)
\end{equation}
with $f(z)$ a holomorphic function that satisfies periodicity conditions provided by the gauge transformation that one needs to apply as one goes around periods of the torus:
\begin{equation}
    f(z + 1) = f(z), \qquad f(z+\tau) = f(z) \exp(-4\pi i q z -2 \pi i q \tau + 2\pi ib)
    \label{eq:qper}
\end{equation}
Any holomorphic function of $z$ obeying \eqref{eq:qper} can be expressed as a product of elliptic theta-functions. Thus, the solutions of \eqref{eq:xi_equation} are
\begin{equation}
\label{eq:full_solution_with_theta_functions}
    \xi_q(z, \bar z) = C \, e^{\frac{\pi ( zq + b )(z - {\zb}) }{\Im \tau}} \, \prod_{i=1}^{2q} \theta_1\left(z - z_i, \tau \right), \end{equation}
where $C$ is some constant factor, and $z_1, \ldots, z_{2q}$ an unordered collection of points on ${\CalE}_{\tau}$ obeying
\beq
\sum_{i=1}^{2q} z_i = b\, . \label{eq:sumb}
\eeq
We thus parametrize the space of relevant deformations as a complex cone (with $C$ parametrizing the cone direction) over  the locus of $Sym^{2q} {\CalE}_{\tau}$ where the sum of these points  in the sense of the abelian group structure of the elliptic curve is fixed for a specific ${\bf A}_0$ by \eqref{eq:sumb}. Taking the union of relevant directions over the space of marginal deformations of ${\bf A}_0$ (parametrized by $b$) produces a cone  over $Sym^{2q} {\CalE}_{\tau}$. 

The space $\CalV$ of functions of the form \eqref{eq:full_solution_with_theta_functions} is a $2q$-dimensional complex vector space\footnote{A basis in that space can be found by representing \eqref{eq:full_solution_with_theta_functions} as:
\beq
{\xi}(z, {\zb}) = \, e^{\frac{\pi ( zq + b )(z - {\zb}) }{\Im \tau}} \, {\theta}_{1}(z-b)\, {\theta}_{1}(z)^{2q-1} \, {\psi}(z)
\label{eq:linear}
\eeq
where $\psi(z) = {\psi}(z+1) = {\psi}(z+{\tau})$ is a meromorphic function on ${\CalE}_{\tau}$ with poles only at $z =b$, of order at most one, and at $z=0$, of order at most $2q-1$. Such functions are rational functions of
$Y,X$, 
\beq
Y = {\wp}^{\prime}(z)\, , \, X = {\wp}(z) = - {\partial}_{z}^{2} {\rm log} {\theta}_{1}(z)
\eeq
which are related by 
\beq
Y^2 = 4 X^3 - g_{2}({\tau}) X - g_{3}({\tau})
\eeq
In terms of $X,Y$ we can write:
\beq
{\psi} = \sum_{i=0}^{q-1} u_{i} X^{i}  + \sum_{i=0}^{q-2} v_{i} Y X^{i} + w \frac{Y + Y_{b}}{X-X_{b}}
\eeq
where $X_{b} = {\wp}(b)\, , \, Y_{b} = {\wp}^{\prime}(b)$. The $q$ parameters $u_i$, $q-1$ parameters $v_i$ and one $w$ provide the linear coordinates on $\CalV$}.

An important property of the solutions that we have found is that, up to a constant factor $C$, the relevant perturbation from the critical point is completely determined by the set of its zeros. The explicit form of the relevant perturbation ${\delta} \bf A$ is, of course, gauge dependent. However, the positions of the zeros are gauge independent. Thus, in any gauge, there is a one-to-one correspondence between the relevant directions near a critical point and sets of $2q$ unordered points.

\subsection{Structure of $\mathcal A(P)$}
\label{sec:structure_of_A(P)}

Now we are ready to assemble the pieces of the space of $SU(2)$-connections $\mathcal A(P)$ in a coherent fashion. A critical point $\mathbb A \in \mathcal M_q$ corresponds to the splitting of the associated rank two vector bundle with a chiral Dirac operator $(\mathcal E, \mathcal D)$ as a direct sum of two $U(1)$ subbundles, $\mathcal E \simeq \mathcal L_q \oplus \mathcal L_{-q}$, $c_{1}(L_{q}) = q$. The operator $\mathcal D$ acts diagonally, i.e. $\mathcal D = \mathcal D_q \oplus \mathcal D_{-q}$. The operator $\mathcal D_q$ has a non-empty kernel only for $q>0$. This is an analogue of the $q$-th Landau level in the integer quantum Hall effect.

Most points in $\mathcal A(P)$ don't lie in a critical subspace. But they flow to such subspaces under the YM flow. Let us denote by ${\mathcal A}_q\subset \mathcal A(P)$ the subset of connections that flow into $\mathcal M_q$. Those subspaces give us a \textit{stratification} of $\mathcal A(P)$, that is, we have
\beq
\mathcal A(P) = \bigcup_{q\ge 0} {\mathcal A}_q, \quad {\overline{{\mathcal A}}}_q\supset {{\mathcal A}}_{q+1}
\eeq
The associated vector bundle $\mathcal E$ is not split in general. Rather, there is a commutative diagram
\begin{equation}
\begin{tikzcd}
	\Gamma(\mathcal L_{q}) \arrow[d, "\mathcal D_{q}"] \arrow[r, "\phi_1"] & \Gamma(\mathcal E) \arrow[d, "\mathcal D"] \arrow[r, "\phi_2"] & \Gamma(\mathcal L_{-q}) \arrow[d, "\mathcal D_{-q}"] \\
	\Gamma(\mathcal L_{q}) \arrow[r, "\phi_1"] & \Gamma(\mathcal E)  \arrow[r, "\phi_2"] & \Gamma(\mathcal L_{-q}) \\
\end{tikzcd}
\end{equation}
where $\phi_1:\mathcal L_q \to \mathcal E$ and $\xi_1:\mathcal E \to \mathcal L_{-q}$ are bundle homomorphisms that satisfy $\phi_2\circ \phi_1 = 0$. The structure of this diagram is such that only the choice of $\mathcal L_q$ in $\mathcal E$ is well-defined. With respect to an arbitrary choice of a complement $\mathcal L_{-q} \subset \mathcal E$ such that $\mathcal E = \mathcal L_{q} \oplus \mathcal L_{-q}$, the operator $\mathcal D$ decomposes as
\beq
\mathcal D = \mathcal D_q \oplus \mathcal D_{-q} + D_{21}
\eeq
where $D_{21}:\mathcal E\to \mathcal E$ is a linear bundle map that satisfies $D_{21}^2=0$, $D_{21}\circ \phi_1 = \phi_2 \circ D_{21} = 0$. The YM flow acts with complexified $SU(2)$ gauge transformations as
\beq
\phi_1 \mapsto g^{-1} \phi_1, \quad \mathcal D \mapsto g^{-1} \mathcal D g, \quad \phi_2 \mapsto g \phi_2
\eeq
This shows that topologically the summands $\mathcal D_{q}$ and $\mathcal D_{-q}$ are not affected by the flow. In addition, as we end up on $\mathcal M_q$, in the $t\to \infty$ limit $D_{12}\to 0$.

We have found above that the space $T_{{\bf A}_{0}}^{+} {\mathcal A}_q$ of relevant deformations form a subspace of the tangent space to $T_{{\bf A}_0} {\mathcal A}_q$ of complex dimension $2q$. In fact, $T_{{\bf A}_{0}}^{+} {\mathcal A}_q$ is the first Dolbeault cohomology group
$$H^{1} (End({\CalE})) = {\Gamma}( {\Omega}^{0,1}_{M} \otimes LieG_{\BC}) / {\rm Im} {\CalD}\ . $$ 
The space $T_{{\bf A}_{0}}^{+} {\mathcal A}_q$ naturally identifies with
the normal bundle to ${\CalA}_{q} \subset {\CalA}(P)$ (we take the fiber over a point close to ${\bf A}_{0}$ and spread it over ${\CalA}_{q}$ by
the action of ${\CalG}_{\BC}$). 
Thus, the real codimension of $\mathcal A_q\in \mathcal A(P)$ is equal to $4q$.

In TBG, however, there is only one parameter $\alpha$. The magic angles should at least correspond to intersections with $\mathcal M_1$, as for $\mathcal D \in \mathcal M_q$ we have $\dim_{\mathbb C} \ker \mathcal D = \dim_{\mathbb C} \ker \mathcal D_q = q$. Generally, unless there is something special about the studied family of connections, we wouldn't find any intersections of that family with $\mathcal M_1$. This motivates the further investigation of $\mathcal A(P)$ in the next section.

\subsection{Stability, zero modes, conformal blocks, and Chern-Simons}

Before getting into the story of invariant connections, let us pause to review yet another point of view on the emergence of flatbands. 

In two dimensional conformal field theory  with $G$-symmetry based current algebra, one studies the generating functions of chiral $G$-currents. Assuming $\Sigma$ is the Riemann surface on which CFT lives, 
\beq
{\bf\Psi} \left[ A_{\bar z} \right] = \left\langle \, {\exp} \int_{\Sigma}
 \, dzd{\bar z}\, {\rm tr} J_{z} A_{\bar z} \right\rangle
\label{eq:currcorr}
\eeq
The standard Ward identities \cite{Gerasimov:1993ws, Witten:1993xi} imply the left hand side of \eqref{eq:currcorr} obey the cocycle condition:
\beq
{\bf\Psi} \left[ g^{-1} A_{\bar z} g + g^{-1} {\bar\partial}_{\bar z} g \right] = e^{k S_{WZW}(g) + k \int_{\Sigma} {\rm tr} ( g^{-1}{\partial}_{z} g A_{\bar z} )} {\bf\Psi} \left[ A_{\bar z} \right]
\label{eq:cocycle}
\eeq
for any $g = g(z, {\bar z}) \in G_{\BC}$. 
Forgetting the origin of the ${\bf\Psi}$-functional, one can take its holomorphy
(perhaps some assumptions on the growth) and the ${\CalG}_{\BC}$-equivariance \eqref{eq:cocycle} as the definition of the level $k \in {\BZ}$ \emph{conformal block} of $\widehat{\mathfrak g}$ current algebra.

Perhaps a useful at this point analogy of the space with the group action and the space of level $k$-equivariant functions is provided by the 
construction of the complex projective space. Consider the vector space ${\mathfrak A} = {\BC}^{M}$, with the action of the group $G = U(1)$
by the simultaneous rotation of the phases of all $M$ coordinates:
\beq
{\bf w} = ( w_{1}, \ldots, w_M ) \mapsto e^{\ii\vartheta}\cdot {\bf w} : = ( e^{\ii\vartheta} w_{1}, \ldots, e^{\ii\vartheta}  w_M ) 
\label{eq:exw}
\eeq
A level $k$ conformal block is a holomorphic function ${\psi}$ of $\bf w$ obeying
\beq
{\psi} ( e^{\ii\vartheta}\cdot {\bf w} ) = e^{\ii k {\vartheta}} {\psi} ({\bf w}) 
\label{eq:excoc}
\eeq
for any $\vartheta$. Of course, the holomorphy of ${\psi}$ implies 
\eqref{eq:excoc} holds for any complex $\vartheta$. Thus, ${\psi}$ is
$G_{\BC} = {\BC}^{\times}$-equivariant. 

Now, algebraic geometers define a notion of \emph{stability} for points in ${\mathfrak A}$ or ${\CalA}(P)$, defined relative to a choice of cocycle, such as in \eqref{eq:excoc} or \eqref{eq:cocycle}, as the existence of such ${\psi}$ or ${\bf\Psi}$, obeying the $G_{\BC}$ or ${\CalG}_{\BC}$-equivariance, such that it does not vanish at $\bf w$ or $A_{\bar z}$, respectively. The geometric invariant theory identifies
the set of stable points with those whose $G_{\BC}$ (${\CalG}_{\BC}$)-orbit intersects the zero locus ${\mu}^{-1}(0)$ of the moment map. In the finite dimensional example the moment map ${\mu}: {\mathfrak A} \to {\BR}$ is 
\beq
{\mu} = | w_{1} |^2 + \ldots + |w_{M}|^2 - r\, , \ r> 0
\label{eq:moment1}
\eeq
while in the infinite dimensional setting ${\mu} :{\CalA}(P) \to {\Omega}^{2}_{M} \otimes ad(P)^{*}$ is given by the curvature of the unitary connection
\beq
{\mu} = F_{A} = {\partial}_{z} A_{\bar z} - {\partial}_{\bar z} A_{z} + [ A_{z}, A_{\bar z}] 
\label{eq:moment2}
\eeq
The Yang-Mills flow we explore in this paper is a particular case of the flow on K{\"a}hler manifold, generated by the gradient ${\nabla} \Vert {\mu} \Vert^2$ of the squared norm of the moment map. So, if the flow reaches the zero locus of $\mu$, the starting point is stable, and a conformal block, non-vanishing at that starting point and, consequently, at any point along the flow, exists. In the finite dimensional example it is all very trivial: the conformal blocks are simply homogeneous degree $k$ polynomials of $(w_1, \ldots, w_M)$. If ${\bf w} = 0$ any such polynomial vanishes. If ${\bf w} \neq 0$ then for $k>0$ it is easy to find a degree $k$ polynomial which does not vanish at $\bf w$ (if $w_i \neq 0$ for some $i= 1, \ldots, M$, then take ${\psi} = w_{i}^{k}$).

In the infinite-dimensional setting of ${\CalA}(P)$ having a conformal block non-vanishing  at $A_{\bar z}$ means the $\CalG_{\BC}$-orbit of $A_{\bar z}$ passes through the locus ${\CalC}_{0}$ of unitary flat $G$-connections. 

In our story we were interested in $A_{\bar z}$'s, whose ${\CalG}_{\BC}$-orbit doesn't reach ${\CalC}_{0}$. By the above reasoning, it means that any positive level $k>0$ conformal block ${\bf\Psi}$ vanishes at such $A_{\bar z}$. It might be interesting to attribute such a property to the underlying conformal field theory. If the underlying CFT is the theory of free fermions, say $k N$ spin $\frac 12$ Dirac chiral fermions
\beq
I = \int_{\Sigma} \sum_{i=1}^{N} \sum_{f=1}^{k} {\tilde\psi}_{i,f} {\bar\partial} {\psi}^{i,f}
\eeq
where we only look at the $SU(N)$ symmetry, generated by
\beq
\left( J_{z} \right)_{i}^{j} = \sum_{f=1}^{k} : {\tilde\psi}_{i,f} {\psi}^{j,f} : 
\eeq
then the vanishing of the expectation value \eqref{eq:currcorr} is equivalent to having the zero modes
\beq
\left( {\delta}^{i}_{j} {\bar\partial} + A^{i}_{j, \bar z} \right) {\psi}^{j,f} = 0 \, , 
\eeq
killing the fermionic path integral, 
\beq
\int \, D{\tilde\psi} D{\psi} \, e^{- \int_{\Sigma} \sum_{i=1}^{N} \sum_{f=1}^{k} {\tilde\psi}_{i,f} \left( {\delta}^{i}_{j} {\bar\partial} + A^{i}_{j, \bar z} \right) {\psi}^{j,f}} = 0 \ . 
\eeq
We thus understand the connection between the flatbands and stability of $A_{\bar z}$ in the sense of geometric invariants theory.

\section{Discrete group actions in the space of connections}

For some values of $\tau$, there are linear subspaces of $\mathcal A(P)$ invariant under the YM flow. Working within such a subspace, we can reduce the number of dimensions and simplify the problem. This is actually the case for the theory of TMG. One way that could occur is when there is a discrete group $\Gamma$ acting on the base $M$\footnote{  
In general, the multiplication
\beq
z \mapsto {\zeta} z
\eeq
is compatible with the periodicity lattice ${\BZ} + {\tau}{\BZ}$ if
multiplication by $\zeta$ preserves the lattice:
\beq
{\zeta} = m_1 + n_1 {\tau}\, , \ {\zeta}{\tau} = m_{2} + n_{2} {\tau}
\eeq
It follows that $\tau$ is a solution of quadratic equation
\beq
-m_2 + (m_1-n_2)  {\tau} + n_1 {\tau}^{2} = 0
\eeq
with integer
coefficients
\beq
{\tau} \in {\mathbb Q} [ \sqrt{D}]\,  , \ D = (m_1-n_2)^2 + 4 n_1 m_2 < 0
\eeq
Such curves ${\CalE}_{\tau}$ are called, not surprisingly, curves with complex multiplication. In general, the group generated by multiplication
by $\zeta$ is not finite (and does not act in a nice way)}. For special $\tau$'s the group $\Gamma$ is finite:
it is $\mathbb Z_4$ for $\tau=e^{i\pi/2}$, $\mathbb Z_3$ or $\mathbb Z_6$ for $\tau=e^{i\pi/3}$. In this case, the symmetry can be lifted to an action on the bundle $P$ and therefore to an action on the space of connections $\mathcal A(P)$. In a chosen trivialization of the bundle, such lift must act on the connections as 
\beq
{\rho}: \mathbf A \mapsto \rho_g(\mathbf A):=g_{\rho}^{-1}\rho^*(\mathbf A)g_{\rho} + g_{\rho}^{-1}\dd g_{\rho}
\label{eq:rotations_lift}
\eeq
where $\rho \in {\Gamma}$ acts on the base ${\BT}^{2}$ and $g_{\rho} = g_{\rho}(x)\in SU(2)$ realizes the action on the fiber. The map ${\rho} \mapsto g_{\rho}(x)$ must obey the cocycle condition:
\beq
g_{\rho_{1}}(x) \cdot g_{\rho_{2}}({\rho}_{1}\cdot x) = g_{\rho_1 \rho_2}(x)
\eeq
The action of $\Gamma$ commutes with the Yang-Mills flow. Therefore, we can study the flows on the subspace ${\CalA}(P)^{\Gamma}$ of  $\Gamma$-invariant gauge fields.

We note in passing that the possible lifts of the action of $\Gamma$
from $M$ to the bundle $P$ are classified by the collections 
\beq
\bigcup_{m \in M} \, {\rho}_{m} 
\eeq
of homomorphisms 
\beq
{\rho}_{m} : Stab_{m} \longrightarrow G
\eeq
of stabilizers of the points of $M$ (most of them are trivial) to the structure group. Two lifts that share the same $\rho_m$ can be transformed one into another with the gauge transformations.
Next, the action of $\Gamma$ on $\mathcal A(P)$ descends to the tangent space $T_{{\bf A}_0} \mathcal A(P)$ near a critical point $\bf A_0$, provided that ${\Gamma}(\bf A_0)=\bf A_0$. The vector space $T_{{\bf A}_0} \mathcal A(P)$ decomposes as a sum of irreducible representations of $\Gamma$, or, in more physical terms, as a sum of subspaces with different $\Gamma$-charges. For the study of Yang-Mills flow on ${\CalA}(P)^{\Gamma}$, only the subspace $T_{{\bf A}_0} \mathcal A(P)^{\Gamma} \subset T_{{\bf A}_0} \mathcal A(P)$ of $\Gamma$-invariant relevant deformations is relevant. Thus, the expected effective codimension of the set of critical values of $\alpha$ is reduced.

In what follows, we give a more detailed study of this splitting of $T_{{\bf A}_0} \mathcal A(P)$ in specific cases of $\tau = \exp( \frac{\pi\ii}{3} )$ and $\tau = \ii$. 

\subsection{Rotational action in the diagonal gauge}

Consider a critical $SU(2)$ connection $\mathbf A\in \mathcal A_q(P)$ in diagonal gauge given by 
\beq
\mathbf A(z, \bar z) = i a\sigma_3,\quad a(z, \bar z)= - \pi  q \frac{\bar z \dd z - z \dd \bar z}{\tau - \bar \tau}
\label{eq:rotationally_invariant_critical_connection}
\eeq
It is equivalent to \eqref{eq:linear_gauge} up to a gauge transformation. It defines a connection on a trivial vector bundle $\mathcal P_{\mathbb C}: \mathbb C\times \mathbb C^2\to \mathbb C$ that is invariant under translations acting on that bundle
\beq
\begin{split}
\mathbf A(z+1, \bar z + 1) = g_a^{-1}\mathbf A(z,\bar z)g_a + g_a^{-1}\dd g_a,\quad g_a(z,\bar z) = \exp(i\pi q\sigma_3)\exp(i\pi q \sigma_3 \frac{z-\bar z}{\tau-\bar \tau}) \\
\mathbf A(z+\tau, \bar z + \bar \tau) = g_b^{-1}\mathbf A(z,\bar z)g_b + g_b^{-1}\dd g_b,\quad g_b(z,\bar z) = \exp(i\pi q\sigma_3)\exp(i\pi q \sigma_3 \frac{\bar \tau z-\tau \bar z}{\tau-\bar \tau})
\end{split}
\label{eq:translation_action_on_connections}
\eeq
Those transformations define a lift of action of the lattice group $\Lambda = \mathbb Z \oplus\tau \mathbb Z $ on the base space onto the bundle. The factor $\exp(i\pi q \sigma_3)$ was added to make the definition consistent with the rotational symmetry below. A connection $\mathbf A$ that commutes with the action of that lift defines a connection on the factorbundle $\mathcal P_{\mathbb C} / \Lambda \simeq\mathcal P$. The study of connections on a torus is equivalent to a study of connections on the complex plane invariant under \eqref{eq:translation_action_on_connections}.

Now, we can try to expand the group acting on the base space. In particular, for values of $\tau=\exp(\pi i/3)$ and $\tau = i$, we can also add a discrete rotational group $\Gamma$ acting by $z\mapsto \omega z$, $\omega\in \Gamma$, expanding the total group action to $\Gamma \ltimes \Lambda$. The expanded action on the base can be lifted onto the bundle in a consistent way. For a concrete example, the connection \eqref{eq:rotationally_invariant_critical_connection} is invariant under both translations \eqref{eq:translation_action_on_connections} and rotations acting by
\beq
\mathbf A(\omega z, \omega^{-1} \bar z) = g_\omega^{-1}\mathbf A(z,\bar z)g_\omega + g_\omega^{-1}\dd g_\omega,\quad g_\omega(z,\bar z) = \diag(\omega^k, \omega^{-k})
\label{eq:rotational_action_in_diagonal_gauge}
\eeq
Together, for any $k\in \mathbb Z$ \eqref{eq:translation_action_on_connections} and \eqref{eq:rotational_action_in_diagonal_gauge} generate an action of $\Gamma \ltimes \Lambda$ on $\mathcal P_\mathbb C$. In particular cases, we will classify the relevant perturbations in $T_{\mathbf A_0}^{+}$, which generally are given by \eqref{eq:full_solution_with_theta_functions},  according to the expanded action that they are preserved by. In practical term, this corresponds to calculating the charges of \eqref{eq:full_solution_with_theta_functions} with respect to rotations.

\subsection{Hexagonal/triangular lattice, ${\Gamma}=\mathbb{Z}_3$}
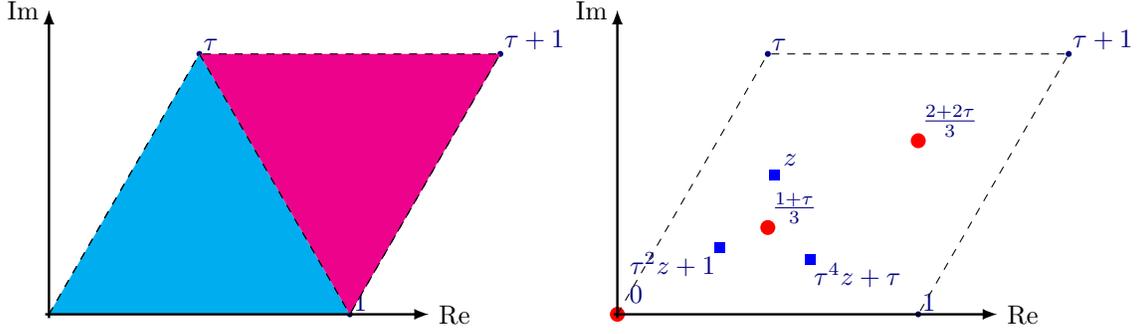
\begin{figure}
\centering
\begin{subfigure}{.5\textwidth}
  \centering
\begin{tikzpicture}
  \def\xmax{5}
  \def\ymax{4}
  \def\R{4}
  \def\angp{83}
  \def\Rp{0.7}
  \def\ang{60}
  \coordinate (O) at (0,0);
  \coordinate (R) at (\ang:\R);
  \coordinate (R2) at (\R,0);
  \coordinate (R+R2) at ($(R)+(R2)$);
  \coordinate (B) at ($1/3*(R)+1/3*(R2)$);
  \coordinate (C) at ($2/3*(R)+2/3*(R2)$);
  \coordinate (D1) at ($(B)+(\angp:\Rp)$);
  \coordinate (D2) at ($(B)+(120+\angp:\Rp)$);
  \coordinate (D3) at ($(B)+(240+\angp:\Rp)$);
  
  \node[fill=mydarkblue,circle,inner sep=0.8] (R') at (R) {};
  \node[mydarkblue,above right=-2] at (R') {$\tau$};
  \node[fill=mydarkblue,circle,inner sep=0.8] (R1') at (R+R2) {};
  \node[mydarkblue,above right=-2] at (R1') {$\tau+1$};
  \node[fill=mydarkblue,circle,inner sep=0.8] (R2') at (R2) {};
  \node[mydarkblue,above right=-2] at (R2') {$1$};
    \draw[fill=cyan, dashed] (O) -- (R) -- (R2);
    \draw[fill=magenta, dashed] (R) -- (R2) -- (R+R2);
  \draw[->,line width=0.9] (-0.05,0) -- (\xmax+0.05,0) node[right] {Re};
  \draw[->,line width=0.9] (0,-0.05) -- (0,\ymax+0.05) node[left] {Im};
  \draw[dashed] (O) -- (R') -- (R1') -- (R2');
\end{tikzpicture}
  \caption{Two halfs of the fundamental region of the torus\newline \newline}
  \label{fig:cyanmagenta}
\end{subfigure}%
\begin{subfigure}{.5\textwidth}
  \centering
\begin{tikzpicture}
  \def\xmax{5}
  \def\ymax{4}
  \def\R{4}
  \def\angp{83}
  \def\Rp{0.7}
  \def\ang{60}
  \coordinate (O) at (0,0);
  \coordinate (R) at (\ang:\R);
  \coordinate (R2) at (\R,0);
  \coordinate (R+R2) at ($(R)+(R2)$);
  \coordinate (B) at ($1/3*(R)+1/3*(R2)$);
  \coordinate (C) at ($2/3*(R)+2/3*(R2)$);
  \coordinate (D1) at ($(B)+(\angp:\Rp)$);
  \coordinate (D2) at ($(B)+(120+\angp:\Rp)$);
  \coordinate (D3) at ($(B)+(240+\angp:\Rp)$);
  
  \node[fill=mydarkblue,circle,inner sep=0.8] (R') at (R) {};
  \node[mydarkblue,above right=-2] at (R') {$\tau$};
  \node[fill=mydarkblue,circle,inner sep=0.8] (R1') at (R+R2) {};
  \node[mydarkblue,above right=-2] at (R1') {$\tau+1$};
  \node[fill=mydarkblue,circle,inner sep=0.8] (R2') at (R2) {};
  \node[mydarkblue,above right=-2] at (R2') {$1$};
  \node[fill=red,circle,inner sep=2] (0') at (O) {};
  \node[mydarkblue,above right=1]  at (0') {$0$};
  \node[fill=red,circle,inner sep=2] (B') at (B) {};
  \node[mydarkblue,above right=-2] at (B') {$\frac{1+\tau}{3}$};
  \node[fill=red,circle,inner sep=2] (C') at (C) {};
  \node[mydarkblue,above right=-2] at (C') {$\frac{2+2\tau}{3}$};
  \node[fill=blue,inner sep=2] (D1') at (D1) {};
  \node[mydarkblue,above right=0] at (D1') {$z$};
  \node[fill=blue,inner sep=2] (D2') at (D2) {};
  \node[mydarkblue,below left=-2] at (D2') {$\tau^2 z + 1$};
  \node[fill=blue,inner sep=2] (D3') at (D3) {};
  \node[mydarkblue,below right=-2] at (D3') {$\tau^4 z + \tau$};

  \draw[->,line width=0.9] (-0.05,0) -- (\xmax+0.05,0) node[right] {Re};
  \draw[->,line width=0.9] (0,-0.05) -- (0,\ymax+0.05) node[left] {Im};
  \draw[dashed] (O) -- (R') -- (R1') -- (R2');
  
\end{tikzpicture}
  \caption{Orbits of $\mathbb Z_3$: one-point orbits are represented by red circles and a three-point orbit is represented by blue squares}
  \label{fig:stable_orbits}
\end{subfigure}
\caption{The case $\tau = e^{\pi\ii /3}$}
\end{figure}

As the simplest example, we will discuss the case of $\tau=\exp(\frac{\ii \pi}{3})$ and the action of $\mathbb Z_3$ on $\mathbb C$ given by $z\mapsto \tau^2 z$. If we consider the action on the torus $\mathbb T^2 = \mathbb C/\Lambda$, it is given by $z \mapsto \tau^2 z + 1$ for points in the lower cyan half of the fundamental region and $z \mapsto \tau^2 z + 2$ for points in the upper magenta half, as shown on Fig. \ref{fig:cyanmagenta}.

As we have seen above, the relevant operators in the $SU(2)$ case are determined by their zeros. Thus, the relevant operators that are invariant under $\rho_g$ for {\it some} $g$ correspond to sets of zeros with multiplicity that are invariant under rotations by $2\pi/3$. The invariant sets of points consist of a union of several orbits of $\mathbb Z_3$. There are three points in a torus that are stable under this action of ${\BZ}_3$, namely $z=0$, $z=\frac{1+{\tau}}{3}$, and $z=\frac{2(1+{\tau})}{3}$. The rest of the orbits consist of triples $\{z, \tau^2 z + 1, \tau^4 z+\tau \}$ with $z$ in the cyan region or $\{z, \tau^2 z + 2, \tau^4 z+2\tau \}$ with $z$ in the magenta region on Fig. \ref{fig:cyanmagenta}. Thus, in the case of the lowest critical point $q=1$, we are only left with a discrete set of relevant perturbations. In the gauge \eqref{eq:linear_gauge} with $b=0$ the connection is trivially symmetric under ${\BZ}_3$. The only symmetric perturbations are
\begin{align}
\label{eq:relevant_symmetric_perturbations_first_level_f1}
    &f(z) = \theta_1^2 \left( z ; \tau \right) \quad \\ 
    \quad &f(z)=\theta_1 \left( z-\frac{1+{\tau}}{3}; \tau \right) \theta_1 \left( z+ \frac{1+\tau}{3} ; {\tau} \right) 
\label{eq:relevant_symmetric_perturbations_first_level_f2}
\end{align}

\begin{figure}
\begin{subfigure}{0.5\textwidth}
    \centering
    \includegraphics[width=\linewidth]{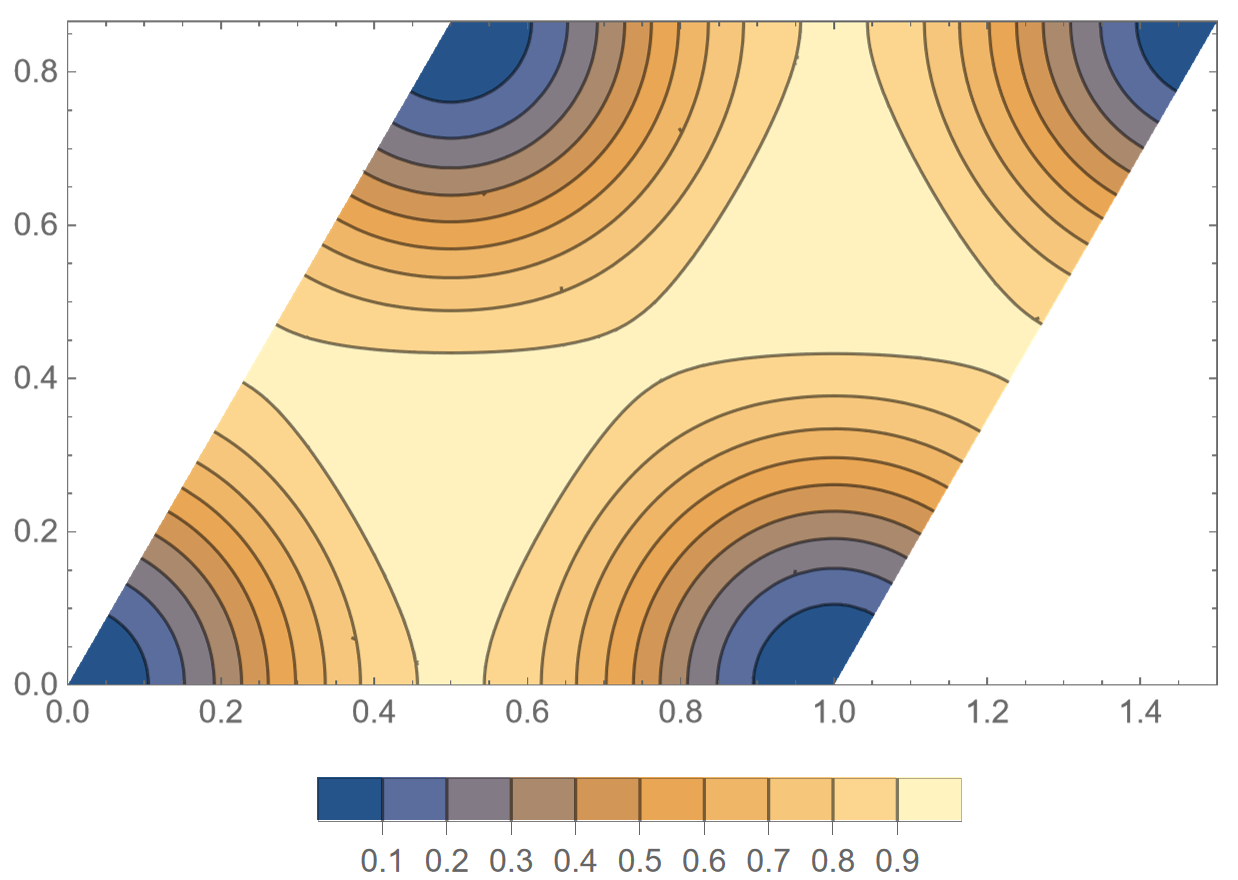}
    \caption{$f(z) = \theta_1^2 \left( z; \tau \right)$ \eqref{eq:relevant_symmetric_perturbations_first_level_f1}}
    \label{fig:relevant_symmetric_perturbation_first_level_f1}
\end{subfigure}
    \begin{subfigure}{0.5\textwidth}
    \centering
    \includegraphics[width=\linewidth]{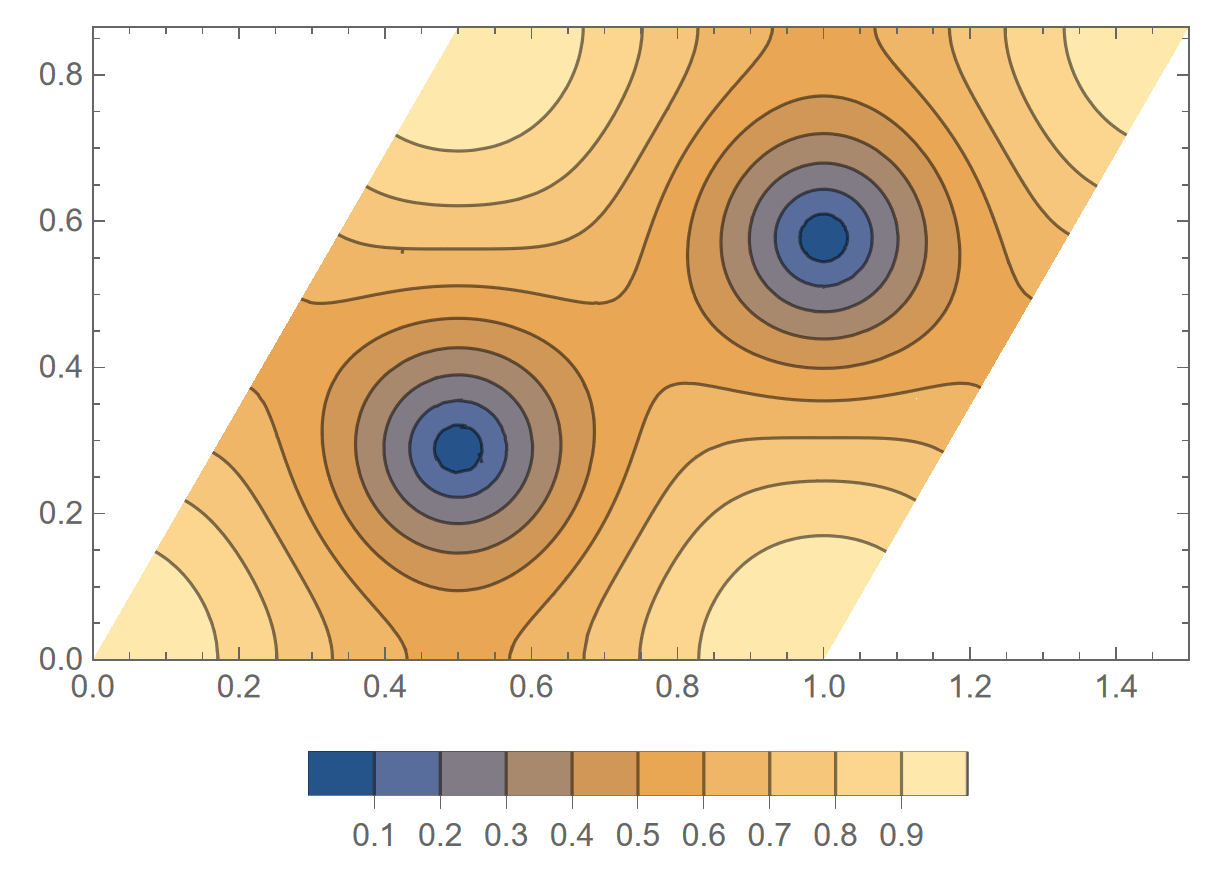}
\caption{$f(z)=\theta_1 \left( z-\frac{1+{\tau}}{3} ; \tau \right)\theta_1 \left( z+\frac{1+{\tau}}{3}; \tau \right)$ \eqref{eq:relevant_symmetric_perturbations_first_level_f2}}
\label{fig:relevant_symmetric_perturbation_first_level_f2}
\end{subfigure}
    \caption{Plots of normalized $|\xi(z,\bar z)|$ for $\tau=e^{\frac{\pi\ii}{3}}$, $q=1$ and symmetric choices of zeros}
    \label{fig:Plots_of_relevant_symmetric_perturbations}
\end{figure}

For higher critical points, there can be one or several triples $\{z, \tau z, \tau^2 z\}$, so the space of relevant symmetric perturbations is at least one-dimensional.

For a generic $q$, any function of the form \eqref{eq:full_solution_with_theta_functions} with a ${\BZ}_{3}$-invariant divisor $Z = ( z_j )$ of zeros is an eigenfunction of the symmetry. Its charge under rotations $z\mapsto \tau^2 z$ is determined by the multiplicity of its zero at $z=0$. E.g. for $q=1$, \eqref{eq:relevant_symmetric_perturbations_first_level_f1} has charge $2$ under rotations, and \eqref{eq:relevant_symmetric_perturbations_first_level_f2} has charge 0.

Thus, the space of relevant directions separates into three distinct subspaces based on the multiplicity of their zeroes at $z=0$. After calculating the dimensions of corresponding function spaces, we get
\begin{equation}
    T_{{\bf A}_0}^{+} \mathcal A(P) \simeq\mathbb C^{2q} = {\BC}^{\floor{\frac{2q+3}{3}}} \otimes {\CalR}_{-q} \oplus {\BC}^{\floor{\frac{2q+1}{3}}} \otimes {\CalR}_{-q+1} \oplus {\BC}^{\floor{\frac{2q-1}{3}}} \otimes {\CalR}_{-q+2}
\end{equation}
where ${\CalR}_{m} \equiv {\CalR}_{m+3}$ is the one-dimensional irreducible representation of ${\BZ}_{3}$, that transforms with the charge $m$ under rotations $\rho_{\bf 1}$ \eqref{eq:rotations_lift}. 

There are three different ways to lift the action of $\mathbb Z_3$ on the torus to its action $\rho_g$ on the bundle leaving the critical point \eqref{eq:rotationally_invariant_critical_connection} invariant. They correspond (in the gauge \eqref{eq:linear_gauge}) to constant matrices $g_m=\exp(\frac{2\pi i m}{3} \sigma_3)$. The space of connections invariant under $\rho_{g_m}$ selects the component of the tangent space with the corresponding charge $m$.

\subsection{Triangular lattice, ${\Gamma}={\BZ}_6 = {\BZ}_3 \times {\BZ}_2$}

\begin{figure}
\centering
\begin{tikzpicture}
  \def\xmax{5}
  \def\ymax{4}
  \def\R{4}
  \def\angp{83}
  \def\Rp{0.7}
  \def\ang{60}
  \coordinate (O) at (0,0);
  \coordinate (R) at (\ang:\R);
  \coordinate (R2) at (\R,0);
  \coordinate (R+R2) at ($(R)+(R2)$);
  \coordinate (B) at ($1/3*(R)+1/3*(R2)$);
  \coordinate (C) at ($2/3*(R)+2/3*(R2)$);
  \coordinate (D1) at ($(B)+(\angp:\Rp)$);
  \coordinate (D2) at ($(B)+(120+\angp:\Rp)$);
  \coordinate (D3) at ($(B)+(240+\angp:\Rp)$);
  \coordinate (D4) at ($(C)+(-\angp:\Rp)$);
  \coordinate (D5) at ($(C)+(120-\angp:\Rp)$);
  \coordinate (D6) at ($(C)+(240-\angp:\Rp)$);
  \coordinate (E1) at ($1/2*(R)$);
  \coordinate (E2) at ($1/2*(R2)$);
  \coordinate (E3) at ($(R)+1/2*(R2)$);
  \coordinate (E4) at ($1/2*(R)+(R2)$);
  \coordinate (E5) at ($1/2*(R)+1/2*(R2)$);

  \node[fill=mydarkblue,circle,inner sep=0.8] (R') at (R) {};
  \node[mydarkblue,above right=-2] at (R') {$\tau$};
  \node[fill=mydarkblue,circle,inner sep=0.8] (R1') at (R+R2) {};
  \node[mydarkblue,above right=-2] at (R1') {$\tau+1$};
  \node[fill=mydarkblue,circle,inner sep=0.8] (R2') at (R2) {};
  \node[mydarkblue,above right=-2] at (R2') {$1$};
  \node[fill=red,circle,inner sep=2] (0') at (O) {};
  \node[mydarkblue,above right=1]  at (0') {$0$};
  \node[fill=orange,circle,inner sep=2] (B') at (B) {};
  \node[fill=orange,circle,inner sep=2] (C') at (C) {};
  \node[fill=blue,inner sep=2] (D1') at (D1) {};
  \node[fill=blue,inner sep=2] (D2') at (D2) {};
  \node[fill=blue,inner sep=2] (D3') at (D3) {};
  \node[fill=blue,inner sep=2] (D4') at (D4) {};
  \node[fill=blue,inner sep=2] (D5') at (D5) {};
  \node[fill=blue,inner sep=2] (D6') at (D6) {};
  \node[fill=green,inner sep=2] (E1') at (E1) {};
  \node[fill=green,inner sep=2] (E2') at (E2) {};
  \node[fill=green,inner sep=2] (E5') at (E5) {};

  \draw[->,line width=0.9] (-0.05,0) -- (\xmax+0.05,0) node[right] {Re};
  \draw[->,line width=0.9] (0,-0.05) -- (0,\ymax+0.05) node[left] {Im};
  \draw[dashed] (O) -- (R') -- (R1') -- (R2');
  \draw[dashed] (R') -- (R2');
  
\end{tikzpicture}
  \caption{Different orbits of $\mathbb Z_6$ in different colors: one each of sizes 1 (red), 2 (orange), 3 (green), 6 (blue)}
  \label{fig:stable_orbits_Z6}
\end{figure}
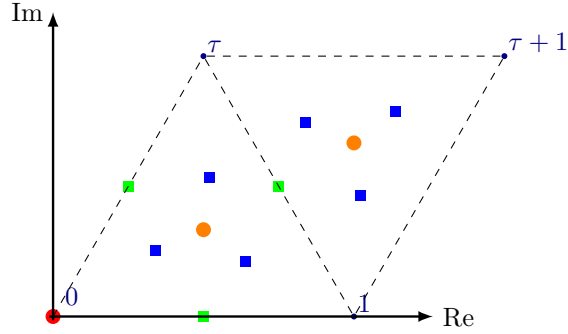

We now consider the case of the torus with $\tau = \exp(\frac{\pi\ii}{3})$. The larger symmetry group $\mathbb{Z}_6$ is generated by 
\beq
z \mapsto {\tau} z  \, {\rm mod}\, {\BZ} + {\tau}{\BZ}
\label{eq:z6ac}
\eeq
One can project the action \eqref{eq:z6ac} onto the fundamental domain: $z\mapsto \tau z + 1$ in the lower blue half of the fundamental region and $z\mapsto \tau z + 1 -\tau$ in the upper magenta half, as shown on Fig. \ref{fig:cyanmagenta}. Examples of different orbits of this action are shown on Fig. \ref{fig:stable_orbits_Z6}.

In a way similar to the $\mathbb{Z}_3$, the space of relevant deformations separates into subspaces labeled by their charges under the $\mathbb{Z}_6$ rotation. An eigendeformation has a divisor of zeroes that is invariant under $\mathbb{Z}_6$. The charge of the deformation is equal to the degree of its zero at $z=0$. Two deformations have the same charge if their sets of zeroes can be continuously deformed one into another without breaking the $\mathbb{Z}_6$ symmetry. In that case, the space of relevant deformations split as
\begin{equation}
    T_{{\bf A}_0}^{+} \mathcal A(P) \simeq\mathbb C^{2q} = \bigotimes_{\substack{n_1\in \{0,1,2\}\\n_2\in \{0,1\}}}{\BC}^{\floor{\frac{2q-2n_1-3n_2}{6}}+1} \otimes {\CalR}_{2q-2n_1-3n_2}
\end{equation}

At particural levels $q\le 4$, the dimensions of the subspaces of different weights are shown in the Table \ref{table:Z6_charges}.

\begin{table}[h]
\begin{center}
\begin{tabular}{|c|cccccc|}
\hline
\multirow{2}{*}{$q$} & \multicolumn{6}{c|}{${\BZ}_6$ charge}                                                                                              \\ \cline{2-7} 
                     & \multicolumn{1}{c|}{0} & \multicolumn{1}{c|}{1} & \multicolumn{1}{c|}{2} & \multicolumn{1}{c|}{3} & \multicolumn{1}{c|}{4} & 5 \\ \hline
1                    & \multicolumn{1}{c|}{1} & \multicolumn{1}{c|}{0} & \multicolumn{1}{c|}{1} & \multicolumn{1}{c|}{0} & \multicolumn{1}{c|}{0} & 0 \\ \hline
2                    & \multicolumn{1}{c|}{1} & \multicolumn{1}{c|}{1} & \multicolumn{1}{c|}{1} & \multicolumn{1}{c|}{0} & \multicolumn{1}{c|}{1} & 0 \\ \hline
3                    & \multicolumn{1}{c|}{2} & \multicolumn{1}{c|}{1} & \multicolumn{1}{c|}{1} & \multicolumn{1}{c|}{1} & \multicolumn{1}{c|}{1} & 0 \\ \hline
4                    & \multicolumn{1}{c|}{2} & \multicolumn{1}{c|}{1} & \multicolumn{1}{c|}{2} & \multicolumn{1}{c|}{1} & \multicolumn{1}{c|}{1} & 1 \\ \hline
\end{tabular}
\end{center}
\caption{Dimensions of subspaces of $T^+_{\mathbf A_0} \mathcal A(P)$ with given charges under ${\BZ}_6$ rotations}
\label{table:Z6_charges}
\end{table}

\subsection{Square lattice,  ${\Gamma}={\BZ}_4$}

\begin{figure}
  \centering
\begin{tikzpicture}
  \def\xmax{5.5}
  \def\ymax{5.5}
  \def\R{5}
  \def\angp{67}
  \def\Rp{1.3}
  \def\ang{90}
  \coordinate (O) at (0,0);
  \coordinate (R) at (\ang:\R);
  \coordinate (R2) at (\R,0);
  \coordinate (R+R2) at ($(R)+(R2)$);
  \coordinate (B) at ($1/2*(R)+1/2*(R2)$);
  \coordinate (D1) at ($(B)+(\angp:\Rp)$);
  \coordinate (D2) at ($(B)+(90+\angp:\Rp)$);
  \coordinate (D3) at ($(B)+(180+\angp:\Rp)$);
  \coordinate (D4) at ($(B)+(270+\angp:\Rp)$);
  
  \node[fill=mydarkblue,circle,inner sep=0.8] (R') at (R) {};
  \node[mydarkblue,above right=-2] at (R') {$i$};
  \node[fill=mydarkblue,circle,inner sep=0.8] (R1') at (R+R2) {};
  \node[mydarkblue,above right=-2] at (R1') {$1+i$};
  \node[fill=mydarkblue,circle,inner sep=0.8] (R2') at (R2) {};
  \node[mydarkblue,above right=-2] at (R2') {$1$};
  \node[fill=red,circle,inner sep=2] (0') at (O) {};
  \node[mydarkblue,above right=1]  at (0') {$0$};
  \node[fill=red,circle,inner sep=2] (B') at (B) {};
  \node[mydarkblue,above right=-2] at (B') {$\frac{1+\tau}{2}$};
  \node[fill=blue,inner sep=2] (D1') at (D1) {};
  \node[mydarkblue,above right=0] at (D1') {$z$};
  \node[fill=blue,inner sep=2] (D2') at (D2) {};
  \node[mydarkblue,below left=-2] at (D2') {$\tau z + 1$};
  \node[fill=blue,inner sep=2] (D3') at (D3) {};
  \node[mydarkblue,below right=-2] at (D3') {$\tau^2 z + \tau + 1$};
  \node[fill=blue,inner sep=2] (D4') at (D4) {};
  \node[mydarkblue,below right=-2] at (D4') {$\tau^3 z + \tau$};

  \draw[->,line width=0.9] (-0.05,0) -- (\xmax+0.05,0) node[right] {Re};
  \draw[->,line width=0.9] (0,-0.05) -- (0,\ymax+0.05) node[left] {Im};
  \draw[dashed] (O) -- (R') -- (R1') -- (R2');
  
\end{tikzpicture}
  \caption{Orbits of $\mathbb Z_4$ for $\tau = i$: one-point orbits are represented by red circles and a four-point orbit is represented by blue squares}
  \label{fig:stable_orbits_square}
\end{figure}
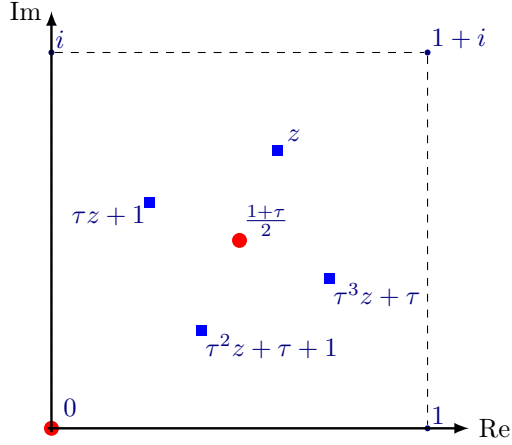

In the case $\tau = \ii$, the fundamental region of the torus is a square. The group $\mathbb Z_4$ acts  by rotations by $\pi/2$. Unlike the triangular case, there are only two fixed points: $z=0$ and $z=(1+{\ii})/2$. The rest of the orbits have order $4$ (Fig. \ref{fig:stable_orbits_square}).

The relevant operators split as
\begin{equation}
     T_{{\bf A}_0}^{+} \mathcal{A}(P) \simeq \bigotimes_{n=0}^3{\BC}^{\floor{\frac{2q-n}{4}}+1} \otimes {\CalR}_{2q-n}
\end{equation}

\subsection{$\Gamma$-invariant flat connections}

The YM flow of a generic connection $A$ on a trivial bundle approaches a flat connection $A_{\infty} \in {\CalC}_{0}$ in the limit $t\to \infty$. Again, the $\Gamma$-invariance of the flow implies that a generic $\Gamma$-invariant YM flow line ends on a $\Gamma$-invariant flat connection. 

The moduli space ${\CalM}_{0}$ of flat $SU(2)$ connections on a torus is a real two dimensional orbifold ${\BT}^2/{\BZ}_{2}$, which has a complex structure, identifying it with ${\BC\BP}^{1}$ with four special points. The $\Gamma$-action on ${\CalA}(P)$, and, consequently, on ${\CalM}_{0}$ is holomorphic, therefore its fixed point locus is zero dimensional. Let us look in detail at the case of $\mathbb Z_6$ action, which will be relevant later for our TBG study.

Let the topological torus ${\BT}^{2}$ be realized as the quotient of ${\BR}^{2}$ by the action of ${\BZ}^2$. The connections on the $SU(2)$ bundle over ${\BR}^{2}$ that correspond to $\Gamma$-invariant connections
(not the gauge equivalence classes!) over the ${\BT}^{2}$ are those that are invariant under the action of $\mathbb Z^2 \ltimes \mathbb Z_6$. A simple calculation shows that the fixed point locus ${\CalM}_{0}^{\Gamma}$ (i.e. gauge equivalence classes of $\Gamma$-invariant flat connections on ${\BT}^{2}$) consists of a single point characterized by having the traces of the monodromies $M_1, M_2 \in SU(2)$ both equal to $\Tr M_1 = \Tr M_2 = {\tau}^{2}+{\tau}^{-2} = -1$ (see Appendix \ref{sec:monodromies_in_the_presence_of_symmetry_derivation} for the derivation).

\subsection{Structure of $\mathcal A(P)$ with a symmetry}

Let us revisit the structure of the space of connections $\mathcal A(P)$ described in Sec. \ref{sec:structure_of_A(P)} in the presence of a discrete symmetry $\Gamma$ acting on it. We have found that the tangent space of relevant operators near a critical point is decomposed into subspaces of different charges under the symmetry. For example, in the case ${\Gamma}= {\BZ}_{3}$ with $\tau = e^{\frac{\pi\ii}{3}}$, for $q=1$, the space ${\BC}^{2}$ of relevant deformations splits as a sum of two one-dimensional subspaces $\mathbb C^1 \oplus \mathbb C^1$. For higher $q$, the space of relevant deformations is split as a sum of three subspaces: for $q=2$, $\mathbb C^4 = \mathbb C^1 \oplus \mathbb C^1 \oplus \mathbb C^2$, for $q=3$,  $\mathbb C^6 = \mathbb C^1 \oplus \mathbb C^2 \oplus \mathbb C^3$, etc. For a given symmetry acting on $\mathcal A(P)$ and a given $\mathbb Z_3$-symmetric realization $\mathbb A$ of a critical connection at level $q$, the relevant perturbations could correspond to any of the 3 possible charges depending on the particular form of $\mathbb A$. 

The diagonal gauge description of critical points and their tangent spaces in this section is parametrized by two numbers: the level $q\in \mathbb Z_{>0}$ and the half-integer charge $k\in \frac{1}{2}\mathbb Z$ defining the action of $\Gamma$ \footnote{Half-integer charge $k$ is allowed because $SU(2)$ acts on the space of connections in the adjoint representation, in which $\text{diag}(-1, -1)$ acts trivially. Therefore, it defines a good action of $\mathbb Z_6$ on the connections. However, from the perspective of its action on the bundle, it corresponds to the action of $\mathbb Z_{12}$ instead of $\mathbb Z_6$.}. One thing left to do is match these data with a generic action of $\Gamma$ on $\mathcal A(\mathcal P)$ defined by the map $\rho$ \eqref{eq:rotations_lift}. This will allow us to establish the possible critical points $\mathbf A_q$ and the dimensions of $T_{\mathbf A_q}^+\mathcal A^\Gamma(\mathcal P)$ present in $\mathcal A^\Gamma(\mathcal P)$. The gauge transformations used to diagonalize the critical point $\mathbf A_q$ do not leave the map $\rho$ invariant. But, as we have noted above, the only invariants of $\rho$ are the conjugacy classes of the action $\rho_m:Stab_m\to G$ in the fibers above base points that have a non-trivial stabilizer subgroup $G\subset \Gamma$.

In the particular case of $\tau = \exp(\frac{i\pi}{3})$ and $\Gamma = \mathbb Z_6$, the action of $\rho_m$ for given $q$ and $k$ is given in Table \ref{table:rho_m}. It will be important for the discussion of symmetries arising in Twisted Bilayer Graphene on the space of connections later in this paper.

\begin{table}[h]
\begin{center}
\begin{tabular}{|c|c|c|}
\hline
Stable point $z_m$    & Stabilizer group $Stab_m$ & Generator of the action on the fiber      \\ \hline
$0$                   & $\mathbb Z_6$             & $\exp(\frac{\ii \pi k}{3}\sigma_3)$         \\ \hline
$\frac{1}{3}(\tau+1)$ & $\mathbb Z_3$             & $\exp(\frac{2\pi\ii}{3} (k+2q)\sigma_3)$ \\ \hline
$\frac{2}{3}(\tau+1)$ & $\mathbb Z_3$             & $\exp(\frac{2\pi\ii}{3} (k+2q)\sigma_3)$ \\ \hline
$\frac{1}{2}$         & $\mathbb Z_2$             & $\exp({\ii} \pi (k+q) \sigma_3)$                  \\ \hline
\end{tabular}
\end{center}
\caption{Diagonalized $\rho_m$ as functions of $q$ and $k$}
\label{table:rho_m}
\end{table}

No matter the complexity of that infinite-dimensional space, it is now clear that the effective codimension of ${\CalA}_q$ is reduced by passing to ${\CalA}_{q}^{\Gamma}$, the subcell of $\Gamma$-invariant connections. Therefore, the appearance of magic angles in graphere is no longer a surprise -- it is to be expected once the chiral splitting is achieved, as explained in \cite{TarnopolskyKruchkovVishwanath}.

\section{Twisted Bilayer Graphene}

Analysis of the critical flow in the presence of a symmetry can be used to explain the appearance of magic angles in the bi-layered graphene. \cite{CaoYuan2018, TarnopolskyKruchkovVishwanath} The phenomenon of magic angles is that for a bilayer graphene structure, at certain relative angles of the layers, a flat electron band appears with zero resistivity. As explained in \cite{TarnopolskyKruchkovVishwanath}, one approximates the one-particle Hamiltonian by the chiral Dirac operator ${\CalD} = {\partial}_{\zb} + A_{\zb}$, with $A_{\zb}$ a $(0,1)$-component of the following connection in an $SU(2)$-bundle
\begin{equation}
\label{eq:graphene_connection}
    A_{\bar z} = \frac{\ii \alpha}{2} \begin{pmatrix}
         0 &  U(\bf r) \\ 
         U(-\bf r) & 0
    \end{pmatrix}, \quad A_{z} = -A_{\bar z}^\dagger
\end{equation}
with\footnote{In rescaled orthonormal coordinates ${\bf r} = (x,y)$ the potential is given by 
\[
U(x, y)=e^{\frac{4\pi i y}{3}}+2e^{-\frac{2i\pi y}{3}} \cos(\frac{2\pi}{3}(\sqrt{3}x-1))\]} \[
U(\bf r) = e^{-\ii\bf q_1 \cdot \bf r} + e^{\ii\phi} e^{-\ii\bf q_2 \cdot \bf r}+e^{-\ii\phi} e^{-\ii\bf q_3 \cdot \bf r}, \] 
with $\phi = \frac{2\pi}{3}$, ${\bf q}_1= \frac{4\pi}{3}(0, -1)$, ${\bf q}_{2,3}=\frac{4\pi}{3}(\pm \sqrt{3}/2, 1/2)$ and $\alpha$ an arbitrary real parameter.

It is quasiperiodic with respect to translations by $\mathbf r_i = \frac{3}{4\pi} \mathbf{q}_i$, each of which changes the phase of $U$ by $e^{i\phi}$. To match the periods of the coordinate $\mathbf r$ to the coordinate $z=\sf x + i \sf y$ on a torus with periods $1, \tau$, we will use the identification $\mathbf r = (\text{Im} z, -\text{Re z})$. The quasiperiodicity allows us to interpret $A_{\bar z}$ as a connection in an $SU(2)$ vector bundle over a torus with the same periods. The band flattens at values of $\alpha$ for which there is a solution of the zero-energy equation $(\bar \partial + A_{\bar z}) \psi = 0$ that has a zero. Since under the Yang-Mills flow the connection $A_{\bar z}$ changes as
\beq
\begin{aligned}
&     \dot A_{\bar z} = -{\ii} \dd_{A_{\bar z}} \star F_{A}  \, , \\
& A_{\bar z}(t) = g^{-1} A_{\bar z}(0) g + g^{-1} {\bar\partial}g \, , \\
& A_{z}(t) = g^{\dagger} A_{z}(0) g^{\dagger -1} + {\partial}g^{\dagger} g^{\dagger -1} \, , \\
& g^{-1}{\dot g} = \star F_{A} \\
\end{aligned}
\eeq
we flow $\psi(z)$ according to $\dot \psi(z) = i\star F_{A} \psi(z)$. Thus, $\psi(z,t)$ remains a solution of $(\bar \partial + A_{\bar z}) \psi = 0$ while its zeros remain intact.

\subsection{Role of symmetry in the appearance of magic angles of different multiplicities}

A YM flow line starting at an arbitrary connection ends on one of the critical manifolds $\mathcal M_q$. For a critical point at the level $q$, the wave function that solves the zero-energy equation  has exactly $q$ zeros. Thus, we expect to have $q$ independent flat bands for any connection that flows to a critical point at the level $q$.

In general, as shown above, the relevant component of the tangent space at a critical point in the moduli space of connections is of complex dimension $2q$. Therefore, one would expect that in general at least $2$ independent complex parameters are necessary for us to find a critical point. However, in TBG flat bands were observed at certain values of only one parameter $\alpha$. The reason for that is that the connection \eqref{eq:graphene_connection} is symmetric under the action of the cyclic group ${\BZ}_6$ that acts on the base space by rotations and whose action on the connections is generated by
\beq
A \mapsto \begin{pmatrix}
    0 & 1 \\ -1 & 0
\end{pmatrix} R_{\pi / 3}^*(A) \begin{pmatrix}
    0 & -1 \\ 1 & 0
\end{pmatrix}
\label{eq:Z6_symmetry_action}
\eeq
with $R^*_{\pi/3}$ is the pullback of the map $R:z\mapsto e^{\ii\pi/3} z$. 
At the level $q=1$, the codimension of $\hat{\mathcal M}_q$ in the space of all connections with a symmetry is equal to $1$, which explains the observation of magic angles in the complex plane parametrized by $\alpha$.

To see which critical points can occur in $\mathcal A_\Gamma$, we need to match this symmetry to the discussion of the previous section. In order to do so, we will find the action of the stabilizer group $Stab_m\subset \Gamma \ltimes \Lambda$ on the fibers above every non-trivial stable point. Denote the genrators of $\Lambda$ as $a$ and $b$ (corresponding to the translations by $\mathbf q_1$ and $\mathbf q_1+\mathbf q_2$) and the generator of $\Gamma=\mathbb Z_6$ as $\gamma$. They satisfy the identities $ab=ba$, $\gamma^6 = 1$, $\gamma a = b\gamma$, and $\gamma b = a^{-1}b \gamma$. In these notations, all of nontrivial stabilizers and the generators of their actions on the fibers over stable points are given in Table \ref{table:TBG_symmetry_action}.

\begin{table}[h]
\begin{center}
\begin{tabular}{|c|c|c|c|}
\hline
Stable point $z_m$    & $Stab_m$ & Generator of $Stab_m\subset \Gamma \ltimes \Lambda$ & Action on the fiber                                             \\ \hline
$z_1=0$                   & $\mathbb Z_6$             & $\gamma$               & $\begin{pmatrix} 0 & 1 \\ -1 & 0 \end{pmatrix}$                 \\ \hline
$z_2=\frac{1}{3}(\tau+1)$ & $\mathbb Z_3$             & $a\gamma^2$            & $\begin{pmatrix} -\omega & 0 \\ 0 & -\omega^{-1} \end{pmatrix}$ \\ \hline
$z_3=\frac{2}{3}(\tau+1)$ & $\mathbb Z_3$             & $a^2\gamma^2$          & $\begin{pmatrix} -\omega^{-1} & 0 \\ 0 & -\omega \end{pmatrix}$ \\ \hline
$z_4=\frac{1}{2}$         & $\mathbb Z_2$             & $a\gamma^3$            & $\begin{pmatrix} 0 & -\omega \\ \omega^{-1} & 0\end{pmatrix}$   \\ \hline
\end{tabular}
\end{center}
\caption{$\mathbb Z_6$ action on the fibers over fixed points}
\label{table:TBG_symmetry_action}
\end{table}

We can deduce from the group action at $z_1$ that we need $[4k]_6 = [0]_6$. It follows that $[k]_6=[\pm\frac{3}{2}]_6$. The action at $z_2$ and $z_3$ gives $[2k+4q]_6=[\pm 1]_6$. $z_4$ gives $[3(k+q)]_6=[\pm \frac{3}{2}]_6$, which is automatically satisfied. The set of equations has solutions for $q = 3n+1$ and $q=3n+2$, $n\in \mathbb Z_{>0}$. The corresponding representations charges of the relevant operators invariant under that symmetry and their dimensions are highlighted in the following table.

\begin{table}[h]
\begin{center}
\begin{tabular}{|c|llllll|}
\hline
\multicolumn{1}{|l|}{}                      & \multicolumn{6}{c|}{${\BZ}_6$ charge $c=2k-1$}                                                                                                                      \\ \cline{2-7} 
\multicolumn{1}{|l|}{\multirow{-2}{*}{$q$}} & \multicolumn{1}{l|}{0} & \multicolumn{1}{l|}{1} & \multicolumn{1}{l|}{2}                         & \multicolumn{1}{l|}{3} & \multicolumn{1}{l|}{4} & 5 \\ \hline
1                                           & \multicolumn{1}{l|}{1} & \multicolumn{1}{l|}{0} & \multicolumn{1}{l|}{\cellcolor[HTML]{C0C0C0}1} & \multicolumn{1}{l|}{0} & \multicolumn{1}{l|}{0} & 0 \\ \hline
2                                           & \multicolumn{1}{l|}{1} & \multicolumn{1}{l|}{1} & \multicolumn{1}{l|}{\cellcolor[HTML]{C0C0C0}1} & \multicolumn{1}{l|}{0} & \multicolumn{1}{l|}{1} & 0 \\ \hline
3                                           & \multicolumn{1}{l|}{2} & \multicolumn{1}{l|}{1} & \multicolumn{1}{l|}{1}                         & \multicolumn{1}{l|}{1} & \multicolumn{1}{l|}{1} & 0 \\ \hline
4                                           & \multicolumn{1}{l|}{2} & \multicolumn{1}{l|}{1} & \multicolumn{1}{l|}{\cellcolor[HTML]{C0C0C0}2} & \multicolumn{1}{l|}{1} & \multicolumn{1}{l|}{1} & 1 \\ \hline
\end{tabular}
\end{center}
\caption{Allowed combinations of $k$ and $q$ for TBG magic points (highlighted) and corresponding dimensions of $T_{\mathbf A_q}^+ \mathcal A^\Gamma$}
\label{table:TBG_codimensions}
\end{table}

The highlighted cells of the Table \ref{table:TBG_codimensions} show the complex codimensions of $\mathcal A_q^\Gamma$ in $\mathcal A^\Gamma$ and indicate to the number of complex parameters that have to be tuned in order to observe the magic angle of corresponding degeneracy $q$ in a generic family of connections. It is clear that it is natural to expect both $q=1$ and $q=2$ critical points in a generic one-parametric family such as \eqref{eq:graphene_connection}. As we see in the next section, this is exactly the case for the TBG connection \eqref{eq:graphene_connection}.

\subsection{Numerical demonstration}

\begin{figure}[h]
    \centering
\begin{subfigure}{0.49\textwidth}
    \includegraphics[width=\linewidth]{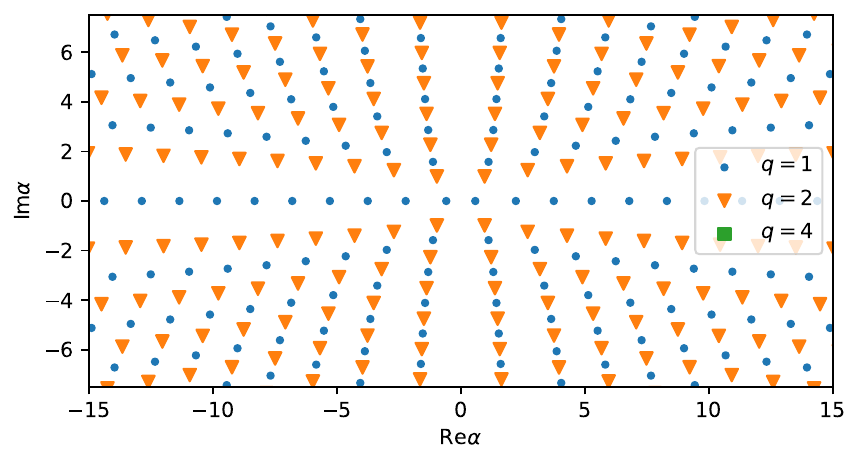}
    \caption{$\beta=0$ (identical to TBG connection)}
    \label{fig:beta_0_magic_angles}
\end{subfigure}
\begin{subfigure}{0.49\textwidth}
    \includegraphics[width=\linewidth]{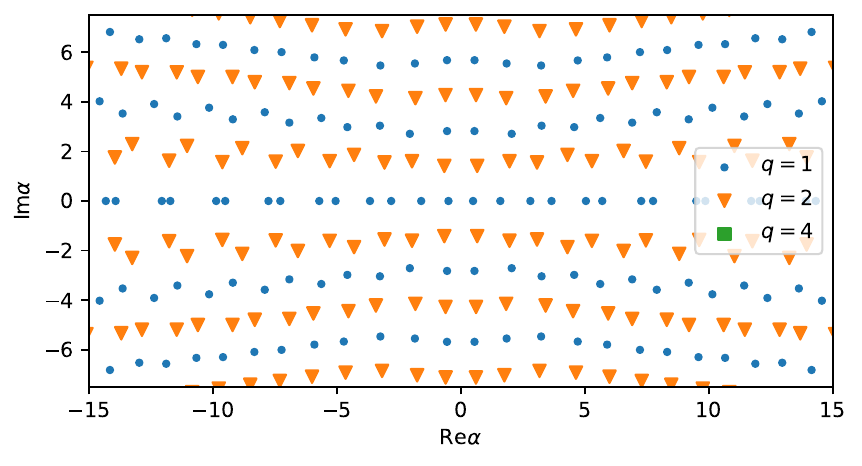}
    \caption{$\beta=0.8$}
    \label{fig:beta_0.8_magic_angles}
\end{subfigure}

\begin{subfigure}{0.49\textwidth}
    \includegraphics[width=\linewidth]{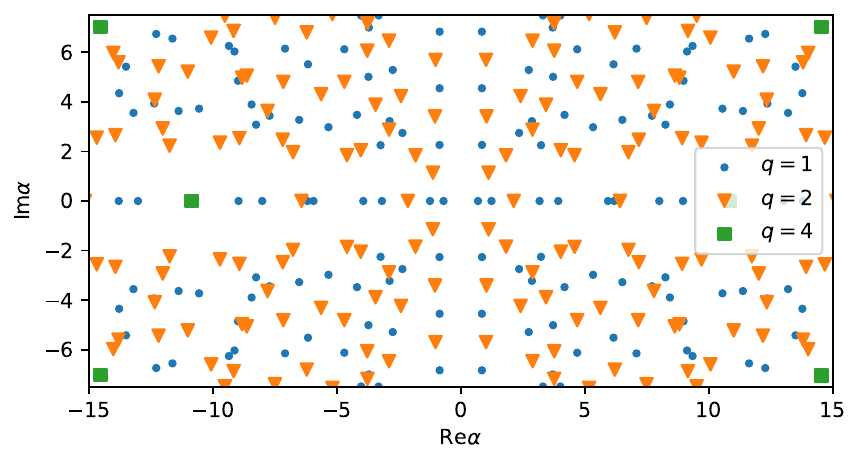}
    \caption{$\beta\approx 1.298$ (notice the appearance of quadruply degenerate flat band)}
    \label{fig:beta_1.298_magic_angles}
\end{subfigure}
    \caption{Values of $\alpha$ corresponding to flat bands of degeneracy $q$ for perturbed connections \eqref{eq:beta_perturbed_connection} for different $\beta$. In other words, values of $\alpha$ for which $\alpha A_{\bar z,(\beta)}$ lies in $\mathcal A_q$.}
\label{fig:magic_angles_plot}
\end{figure}

On the numerical side, we do two types of computer calculations to check our theory. First, we note that the (complex) magic angles $\alpha$ can be found as reciprocals of the eigenvalues of the operator $-\bar \partial^{-1} \circ A_{\bar z}$. The multiplicities of the eigenvalues correspond to the levels $q$ of the critical connections at a given value of $\alpha$. We use the original connection \eqref{eq:graphene_connection}, as well as a more general connection that includes the next harmonic that is symmetric under $\gamma$
\beq
\begin{split}
A_{\bar z,(\beta)} = & \frac{i}{2} \begin{pmatrix}
    0 & U_{(\beta)}(\mathbf r) \\ U_{(\beta)}(-\mathbf r) & 0 
\end{pmatrix}, \\ U_{(\beta)}(\mathbf r) =& \cos(\beta)( e^{-\ii\mathbf q_1 \cdot \mathbf r} + e^{\ii\phi} e^{-i\mathbf q_2 \cdot \mathbf r}+ e^{-\ii\phi} e^{-i\mathbf q_3 \cdot \mathbf r} )+\sin(\beta)( e^{2\ii\mathbf q_1 \cdot \mathbf r} + e^{\ii\phi} e^{2i\mathbf q_2 \cdot \mathbf r}+ e^{-\ii\phi} e^{2i\mathbf q_3 \cdot \mathbf r} )
\end{split}
\label{eq:beta_perturbed_connection}
\eeq
The plots of critical points are shown below. The numerical calculations here align completely with our discussion in previous section. As expected from the codimension argument, at a generic value of $\beta$, both $q=1$ and $q=2$ magic angles are observed (see Fig. \ref{fig:beta_0_magic_angles}, \ref{fig:beta_0.8_magic_angles}). At specific values of $\beta$ (which corresponds to tuning two parameters instead of one), a number of eigenvalues combine into $q=4$ magic angle (Fig. \ref{fig:beta_1.298_magic_angles}).

\begin{figure}[h]
    \centering
\begin{subfigure}{0.46\textwidth}
    \includegraphics[width=\linewidth]{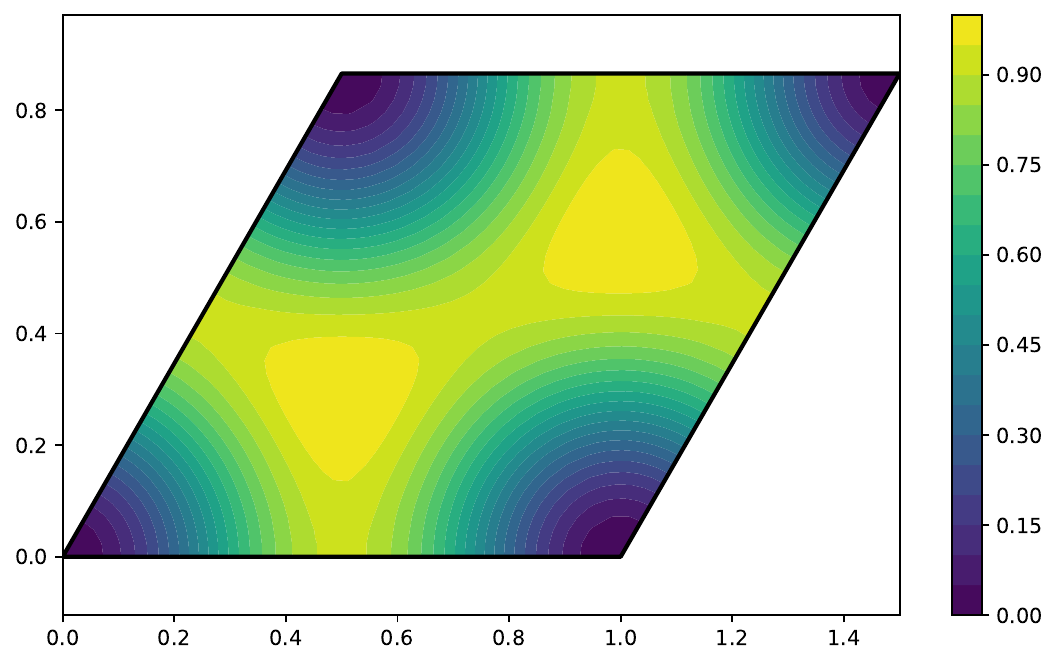}
    \caption{Relevant perturbation $|\xi_1(z,\bar z)|$ consistent with $\mathbb Z_6$ symmetry of TBG.}
\end{subfigure}
\begin{subfigure}{0.46\textwidth}
    \includegraphics[width=\linewidth]{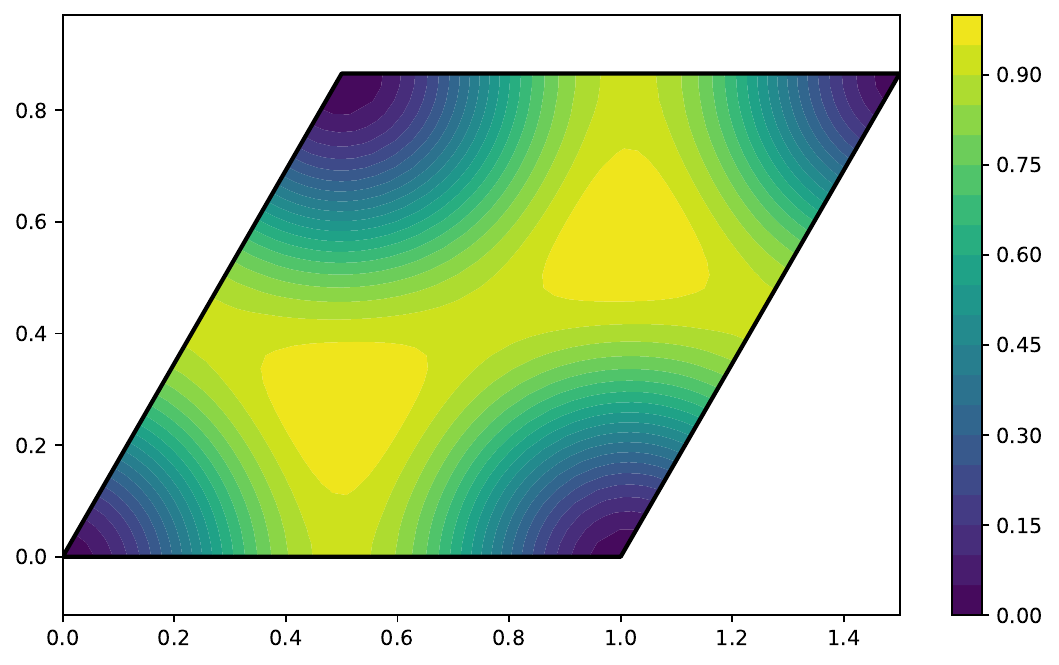}
    \caption{Numerically calculated $|\dot {\mathbf A}|$ near a $q=1$ critical point (starting with $\alpha = 0.600$).}
\end{subfigure}
    
\begin{subfigure}{0.46\textwidth}
    \includegraphics[width=\linewidth]{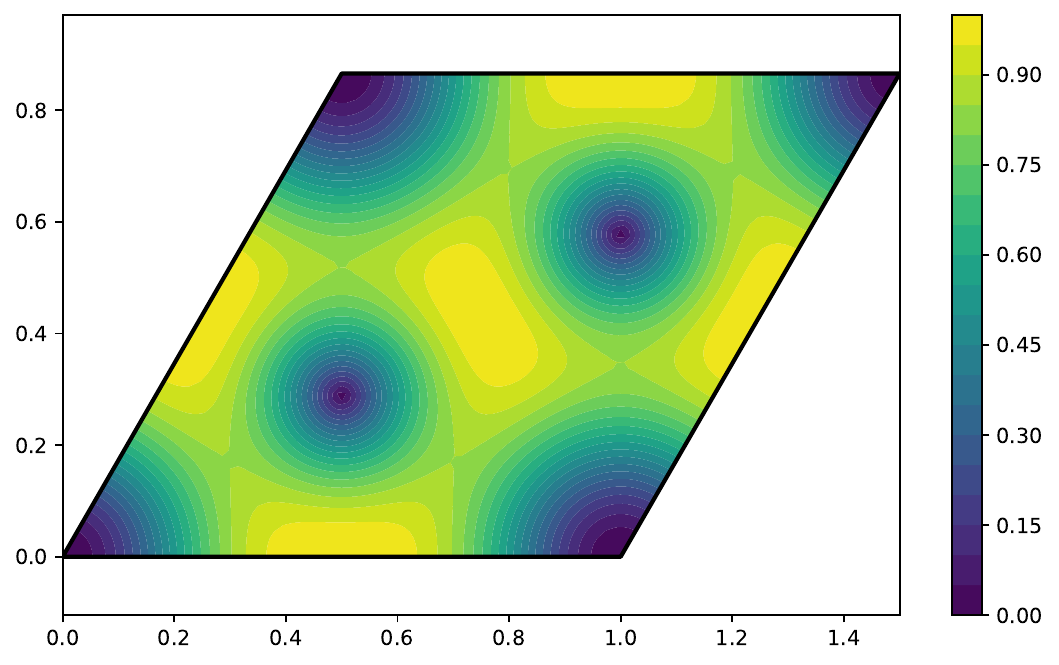}
    \caption{Relevant perturbation $|\xi_2(z,\bar z)|$ consistent with $\mathbb Z_6$ symmetry of TBG at $q=2$.}
\end{subfigure}
\begin{subfigure}{0.46\textwidth}
    \includegraphics[width=\linewidth]{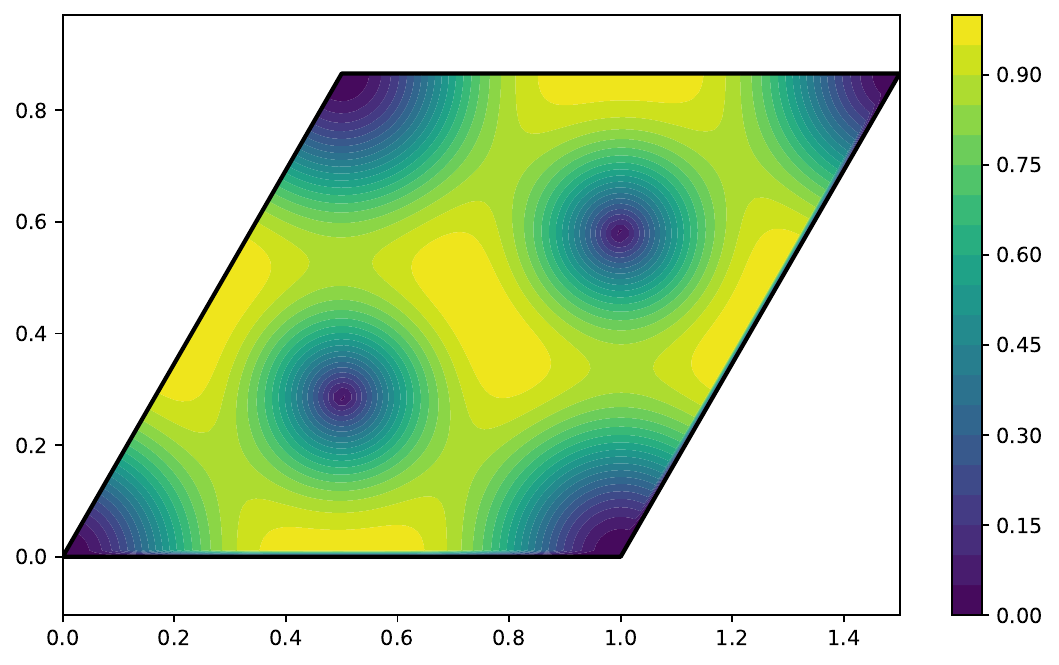}
    \caption{Numerically calculated $|\dot {\mathbf A}|$ near a $q=2$ critical point (starting with $\alpha = 0.96+0.97\ii$).}
\end{subfigure}
    \caption{Absolute values of the relevant perturbations $|{\xi}_q(z,\bar z)|$ \eqref{eq:full_solution_with_theta_functions} and numerical calculations for $|\dot {\mathbf A}|$ when passing the corresponding critical point. The divisors of zeroes of theoretically and numerically found relevant perturbations are clearly the same.}
    \label{fig:relevant_perturbation_plots}
\end{figure}

For an additional check of the theoretical results, we have simulated the YM flow numerically to ensure that it indeed leads to critical points at levels $q=1$ and $q=2$ when starting from a magic angle connection. We also checked that in a neighborhood of magic angles $\dot {\mathbf A}$ corresponds to the relevant perturbations theoretically discussed in Sec. {\bf 3} and {\bf 4}, see Fig. \ref{fig:relevant_perturbation_plots}.

\subsection{Complex $\alpha$, winding number and behavior under perturbations}

By formally treating $\alpha$ as a complex parameter, we might find some additional information about the structure of magic angles. Evaluation of the YM flow in the complex $\alpha$ space reveals that the magic angles are isolated special points. 

 Consider the YM flow at a non-critical point. Due to the preserved $\mathbb{Z}_6$ symmetry, the final flat connection has the monodromy around the a-cycle of the form (See Appendix \ref{sec:monodromies_in_the_presence_of_symmetry_derivation} for derivation)
\begin{equation}
	M_a = \begin{pmatrix}
		-\frac{1}{2} + \frac{\sqrt{3}}{2} i \cos(\phi) & \frac{\sqrt{3}}{2} i \sin(\phi) \\
		\frac{\sqrt{3}}{2}i\sin(\phi) & -\frac{1}{2} - i \frac{\sqrt{3}}{2} \cos(\phi)
	\end{pmatrix}
    \label{eq:monodromy_Z6_form}
\end{equation}

\begin{figure}
    \centering
    \includegraphics[width=0.7\linewidth]{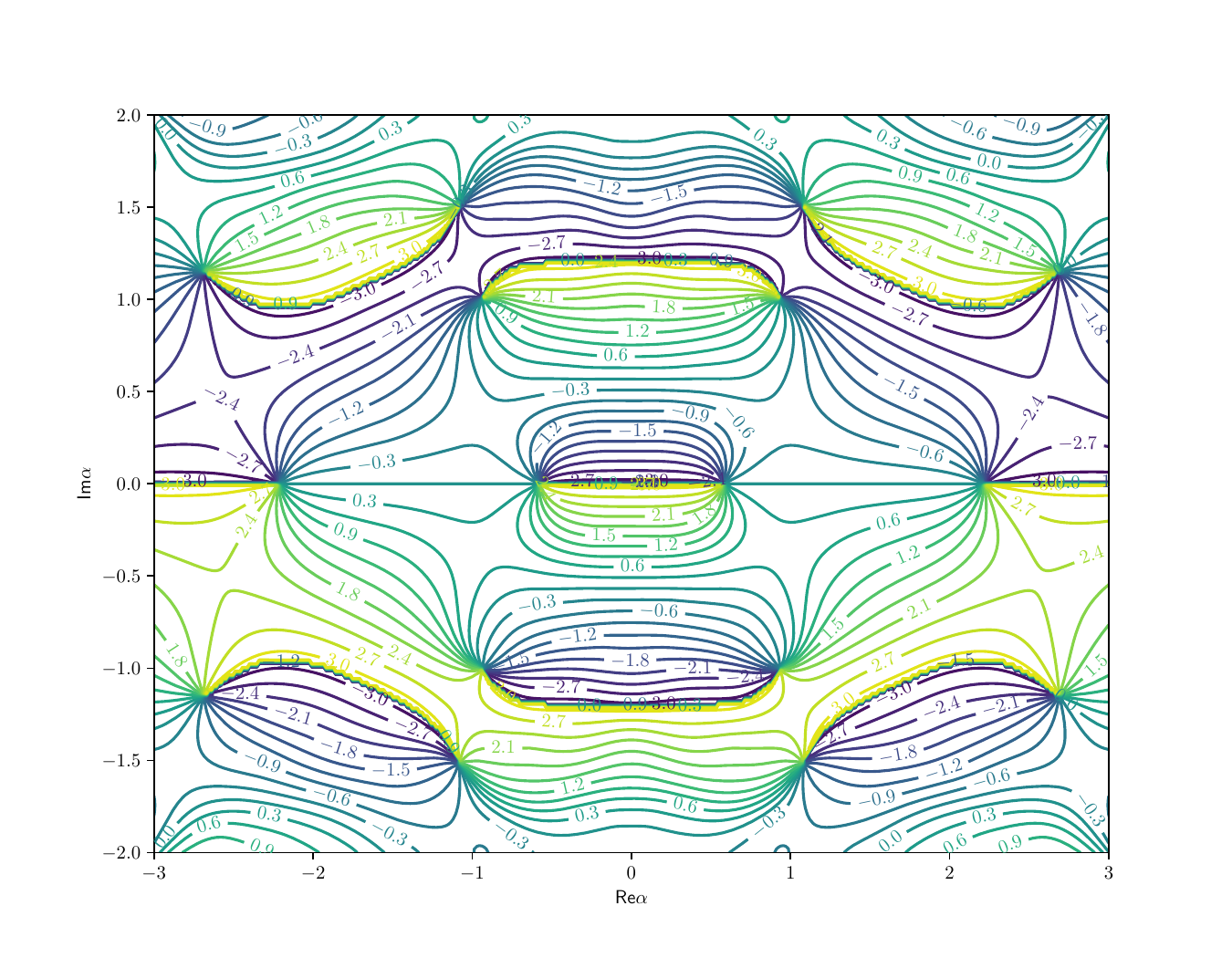}
    \caption{Connection a-cycle monodromy phase $\phi$ (as indicated on the curved lines) at $t\to \infty$ (compare to \cite{Becker2021}, Fig. 2)}
    \label{fig:phases_of_monodromy}
\end{figure}
Therefore, the Yang-Mills flow defines a map $\kappa:\mathbb C\textbackslash \{\pm \alpha_1, \pm \alpha_2, \dots\} \to S^1$. The numerical evaluation shows that the winding number of that map when traveling around a critical point in the $\alpha$ space is $\pm 1$. The behavior of this map for several lower critical points is shown in Fig. \ref{fig:phases_of_monodromy}.

The reason for this behavior of monodromy is that the YM flow commutes with the action of the gauge group $\mathcal G$. In a diagonal gauge near a magic angle $\alpha_i$, up to irrelevant corrections, the connection can be written as
\beq
A_{\bar z}(\alpha_i + \delta \alpha) = \begin{pmatrix}
    a_{\bar z} & 0 \\   {\delta\alpha}\,  \xi & -a_{\bar z}
\end{pmatrix}
\eeq
with $a_{\bar z}$ given by the Eq. \eqref{eq:convenient_gauge_arbitrary_tau}, $\xi = \xi (z, {\bar z})$ given by the Eq. \eqref{eq:full_solution_with_theta_functions}. Note that the phase rotations of the prefactor $C$ in $\xi$ (cf.\eqref{eq:full_solution_with_theta_functions}) are equivalent to gauge transformations with the constant $g = e^{\ii u {\sigma}_{3}}$. Gauge transformations commute with the YM flow; therefore, the monodromies of the final flat connections are transformed as $M_a \mapsto g(u) M_a g(u)^{-1}$ as we go around a critical point. This explains the winding number $|w|=1$ of the monodromy as we go around a critical point.

Now, consider instead of \eqref{eq:graphene_connection} a perturbed connection of the form
\beq
\tilde A_{\bar z} (\alpha, \varepsilon) = A_{\bar z}(\alpha) + \varepsilon \delta A_{\bar z}(\alpha) 
\eeq
If the perturbation $\delta {\bf A}$ also has the abovementioned ${\BZ}_6$ symmetry then the new connection $\tilde A_{\bar z} (\alpha, \varepsilon)$ also produces a map $\kappa(\varepsilon)$ that describes phases $\phi$ of monodromy of the perturbed connection. Due to the continuity of the YM flow away from the critical points, the winding number of the map $\kappa$ around a cycle containing one of the critical points is not changed for sufficiently small $\varepsilon$. Therefore, under such a perturbation, for small $\varepsilon$ the perturbation produces continuous shift $\alpha_i(\varepsilon)$ of magic angles.

For the physical angles $\alpha_i \in \mathbb R$ to remain physical under a perturbation, we need an additional reflection symmetry. The original connection \eqref{eq:graphene_connection} has such a symmetry given by
\beq
\begin{split}
\alpha \mapsto \bar \alpha \\
z \mapsto \bar z \\
A_{\bar z} \mapsto -A_{\bar z}^{\dagger}    
\end{split}
\eeq

This symmetry ensures that if $\alpha$ is a magic angle, then so is $\bar \alpha$.
If the perturbation, as a function of $\alpha$, also has this symmetry, then we are guaranteed that the physical magic angles remain physical, as otherwise they would have to double and the winding number of the map $\kappa$ will change.

A similar result was mentioned in \cite{Becker2021} achieved using the theory of Birman–Schwinger operators.

\subsection{Nontrivial Chern class of the flat band}
In Section \ref{sec:structure_of_A(P)} how certain values of the parameter $\alpha$ lead to the splitting of a trivial $SU(2)$ bundle into a sum of two subbundles with opposite Chern classes $c_1 = \pm 1$. This bundle is defined over a torus in the coordinate space. However, one can show that the Chern class of the $U(1)$ bundle in the momentum space with Berry connection is also nontrivial, which can explain the observation of quantum Hall effect in twisted bilayer graphene at zero magnetic field \cite{Thouless82, NiuThouless85}.

To do so, we first need to find the zero-energy wave function $\psi_{\mathbf k}(\mathbf r)$ at an arbitrary point $\mathbf k$ in the first Brillouin zone $\mathcal B$. The wave function is a solution of the equation
\beq
    (\partial_{\bar z} + A_{\bar z}(\mathbf r)) \psi_{\mathbf k} (\mathbf r) = 0, \quad \psi_k(\mathbf r+\mathbf q_j) = e^{{\ii} \mathbf q_j \cdot \mathbf k}\psi_k (\mathbf r) \, , \ j = 1, 2 \ .
\eeq  
Taking the wave function $\psi_0$ of this equation for $\mathbf k=0$, we can construct a solution for an arbitrary $\mathbf k$. Assuming that $\psi_0(0) = 0$, and working in coordinates where $\mathbf q_1 = (1, 0)$, $\mathbf q_2 = (-\frac{1}{2}, \frac{\sqrt{3}}{2}) $, we have
\beq
\label{eq:zeroshift}
\psi_{\mathbf k} (z, \bar z)= C(\mathbf k)\frac{\theta_1 \left(z-\frac{1}{2\pi \ii}\frac{\sqrt{3}}{2}\left( k_x + {\ii}k_{y} \right); \tau \right)}{\theta_1(z; \tau)}e^{{\ii}k_x z} \psi_0(z, \bar z)
\eeq
where $C(\mathbf k)\in \mathbb R_+$ is a normalizing coefficient such that $\braket{\psi_{\bf k}}=1$. 

The Berry connection is given by
\beq
A_B(k) = \sum_{j= x, y} \bra{\psi_{\mathbf k}} \mathbf \partial_{k_j} \ket{\psi_{\mathbf k}}\dd k_j
\eeq
The form $\frac{1}{2\pi\ii}F_{B}$, with $F_B = dA_B$ represents the first Chern class of the flat band over the Brillouin zone. The latter can also be computed from the transition functions, i.e. by the transformation properties of the wave function under the translations of the momentum $\mathbf k \mapsto \mathbf k + \mathbf k_i$ by the reciprocal lattice vectors: $\mathbf k_1 = 2\pi (0,2/\sqrt{3})$ and $\mathbf k_2 = 2\pi (1, 1/\sqrt{3})$. Explicit calculations give
\beq
\psi_{\mathbf k+ \mathbf k_1}(\mathbf r) = \psi_{\mathbf k}(\mathbf r), \quad \psi_{\mathbf k + \mathbf k_2} =-{\ii}\exp(
2\pi \ii \frac{\mathbf k_1 \cdot \mathbf k}{\mathbf k_1 \cdot \mathbf k_1}) \psi_{\mathbf k} (\mathbf r)
\eeq 

The first Chern class is equal to
\beq
c_1 = \frac{1}{2\pi\ii} \int_{\mathcal B} F_B = \frac{1}{2\pi\ii}\int_{\partial \mathcal B} A_B = 1
\eeq
where we used the fact that $A_B$, being a connection on a $U(1)$ bundle, transforms as $A_B(\mathbf k + \mathbf k_i)=A_B(\mathbf k) - g_i(\mathbf k)^{-1} \dd g_i(\mathbf k)$ if $\psi(\mathbf k+\mathbf k_i)=g_i(\mathbf k) \psi(\mathbf k)$.

\section{Conclusions and further directions}

In this paper, we have related the magic angles in twisted bilayer graphene to topology structure of the space of gauge orbits in the space of gauge fields on a torus in the presence of a discrete symmetry. 
For the TMG the magic angles are identified with the parameters of an $SU(N)$
gauge field $A$, emerging from the structure of the one-particle Hamiltonian in the chiral limit studied, for $N=2$, in \cite{TarnopolskyKruchkovVishwanath}, 
for which the complexified gauge orbit passes through the so-called Yang-Mills connection corresponding to the higher critical point of Yang-Mills action
\beq
S_{\rm YM} = \int_{{\BT}^{2}} {\rm Tr} F_{A} \wedge \star F_{A}
\eeq
Theses critical points can be explicitly classified, and \cite{AtiyahBott} have used them to stratify the space ${\CalA}/{\CalG}$ of gauge equivalence classes and derive powerful results on the topology of the moduli space
${\CalM}_{0} \subset {\CalA}/{\CalG}$ of flat connections (actually, the main interest of \cite{AtiyahBott} was in gauge theory on higher genus Riemann surfaces). 

We work on the space of all gauge fields, as the $SU(N)$-gauge transformations affect the physics of graphene (after all, the gauge fields we are studying are not the dynamical mediators of the electroweak or 
strong force, these are emergent background gauge fields).

In our work we studied the gradient flow on the space $\CalA$ of all gauge fields, driven by $S_{\rm YM}$, 
\beq
{\dot A} = - {\nabla} S_{\rm YM}\ .
\label{eq:ymflow}
\eeq
In these concluding remarks we would like to mention the possible connection of the YM flows to the renormalization group flow. 
The relevant/marginal/irrelevant deformations we discussed are associated with the relevant/marginal/irrelevant deformations of the theory, which we now review. 

Consider two dimensional sigma model with the target space ${\BT}^{2}$, 
\beq
I = \frac{1}{4\pi\alpha'} \int_{\Sigma} (g_{ab} + 2\pi \alpha' {\ii} B_{ab} ) {\partial} X^{a} {\bar\partial} X^{b} 
\label{eq:bulkcft}\eeq
with constant symmetric and antisymmetric $2 \times 2$ matrices $g$ and $B$, respectively. Unitarity requries $g$ be positive-definite. Obviously, the action \eqref{eq:bulkcft} defines a conformal field theory. The stress-tensor is given by:
\beq
T = g_{ab} :{\partial}X^a {\partial} X^b :
\eeq
and obeys the Virasoro OPE, with $c =2$. 
Of course, the chiral algebra of this theory is larger, it contains the $U(1) \times U(1)$ current algebra, generated by
\beq
J^a = {\partial} X^a
\label{eq:currents}
\eeq
Let us now assume $\Sigma$ has a boundary ${\partial}\Sigma$, and let us ask what boundary conditions/boundary interactions are compatible with the conformal symmetry. As far as the boundary conditions are concerned we require
\beq
T = {\bar T}
\eeq
on the boundary, meaning there is no energy flow in or out of the system. For example, the Neumann boundary conditions
\beq
{\partial}_{n} X^a = 0
\label{eq:neumann}
\eeq
are always conformal. In string theory language (even though strictly speaking our sigma model needs additional fields to qualify for a string background) this describes a $D$-brane wrapping the torus ${\BT}^2$. The general background, associated with such a $D$-brane, consists of a gauge field $\bf A$ on the worldvolume of the brane. The background gauge field modifies the action \eqref{eq:bulkcft} by adding a boundary term
\beq
I_{boundary} = \int_{\partial \Sigma} A_{a}(X) dX^a
\label{eq:boundary}
\eeq
Classically, \eqref{eq:boundary} does not use any metric on $\Sigma$, so this term does not violate conformal symmetry. 

Now let us add some quantum mechanical degrees of freedom living only on the boundary (Chan-Paton factors), so that their Hilbert space
is ${\BC}^{N}$ (for $N$-layered graphene). We introduce the $SU(N)$ gauge field ${\bf A} = {\bf A}_{a}(X)dX^a$, with $N \times N$ traceless
antihermitian matrices ${\bf A}_{a}$, $a = 1,2$, and generalize the amplitude $e^{-I_{boundary}}$ to
\begin{multline}
{\rm Tr}_{{\BC}^{N}} \, P {\exp} \, \oint : {\bf A}(X(t)) : = 
N + \frac 12 \int dt_1 dt_2 {\rm Tr} {\CalO}(t_1) {\CalO} (t_2) + \ldots \\
+ \frac{1}{k} \int_{t_1 < t_2 < \ldots < t_{k} < 2{\pi} + t_{1}} 
dt_1 dt_2 \ldots dt_k \ {\rm Tr} {\CalO} (t_1){\CalO}(t_2) \ldots {\CalO}(t_k)+ \ldots  \ , 
\label{eq:pexp}
\end{multline}
where we introduced the normal ordered $N \times N$-matrix valued operators
\beq
{\CalO}(t) = : {\bf A}_{a}(x_0 + {\delta X}(t)) {\partial}_{t}{\delta} X^{a} (t) : 
\eeq

Quantum mechanically the integrand in \eqref{eq:boundary} is singular, since \cite{SW}
\beq
\langle {\delta}X^{a}(t) \, {\delta}X^{b}(t') \rangle =  -{\alpha'} G^{ab} {\rm log} (t - t')^2 + \frac{\ii}{2} {\theta}^{ab} {\epsilon}(t - t'), 
\label{eq:boundprop}
\eeq
for $t \to t'$. 
Hence, the operator \eqref{eq:boundary} needs a regularization, and renormalization. In the first order in ${\alpha}'$-dependence, the dependence of the boundary couplings encoded in $A_{a}(X)$ on the emerging scale $\mu$ will be described by the YM flow \eqref{eq:ymflow}. Details will be published elsewhere. 

We would like to suggest that the conformal field theory given by \eqref{eq:bulkcft} might be of some use for the description of graphene physics.

The YM flow used in our work is a way of stratifying the space of all gauge fields on ${\BT}^2$. The cells corresponding to the YM fields such that the
associated chiral Dirac operator admits a flatband have the codimension
equal to the dimension of the space of relevant perturbations at the
associated fixed point of the YM flow. 
The effective codimension of the set of critical values of the parameter $\alpha$ is reduced for the class of symmetric connections. The study of
$\Gamma$-equivariant YM flows we undertaken in this paper deserves further
mathematical analysis. We found an interesting topology in the neighborhoods of the fixed points, Fig. \ref{fig:phases_of_monodromy}.  It may even play a role in some ramifications
of geometric Langlands program \cite{Frenkel:2009ra}.

\bibliographystyle{plain}
\bibliography{bibliography}

\appendix
\section{Comments on numerical simulation}
For Figures \ref{fig:relevant_perturbation_plots} and \ref{fig:phases_of_monodromy} we simulated the Yang-Mills flow directly using a Python algrorithm solving the PDE on a discretized spacetime lattice. Figure \ref{fig:magic_angles_plot} was obtained as a set of inverse eigenvalues of the operator $\bar \partial^{-1} \circ \mathcal A_{\bar z}$ in the momentum space using numpy.linalg.eig function.

\section{Monodromies of a flat connections on $\mathbb T^2$ with rotational ${\BZ}_3$ and ${\BZ}_6$ symmetries}

\label{sec:monodromies_in_the_presence_of_symmetry_derivation}
Let $E={\BC}^2 \times \mathbb C\to \mathbb C$ be a trivial rank two complex vector bundle over the complex plane. The bundle over a torus can be realized as a factor of $E$ by the action of the group ${\BZ}^{2}$ of lattice translations, covering the action on $\BC$ by shifts from $\Lambda = {\BZ} + \omega {\BZ}$, where $\omega = \exp(2\pi i/3)$. In the case of TBG studied in this paper, translations $z \mapsto z+1$ and $z\mapsto z+\omega$ act on $E$ by simultaneous multiplication of the fibers by $T = \diag(\omega^{-1}, \omega)$. A connection on the complex plane descends to a connection on a torus if it is compatible with the action of translational symmetry. For a flat connection, the monodromies of the connection around the $a-$ and $b-$ cycles of the torus are equal to its monodromies along $[0, 1]$ and $[0, \omega]$ multiplied by $T$.

\begin{figure}[h]
\centering
\begin{tikzpicture}
  \def\xmax{5}
  \def\ymax{5}
  \def\R{4}
  \def\angp{83}
  \def\Rp{0.7}
  \def\ang{60}
  \coordinate (O) at (0,0);
  \coordinate (R0) at (0:\R);
  \coordinate (R1) at (\ang:\R);
  \coordinate (R2) at (2*\ang: \R);
  \coordinate (R3) at (3*\ang: \R);
  \coordinate (R4) at (4*\ang: \R);
  \coordinate (R5) at (5*\ang: \R);
  
  \coordinate (B0) at ($1/3*(R0)+1/3*(R1)$);
  \coordinate (B1) at ($1/3*(R2)+1/3*(R3)$);
  \coordinate (B2) at ($1/3*(R4)+1/3*(R5)$);
  
  \node[fill=mydarkblue,circle,inner sep=0.8] (R0') at (R0) {};
  \node[mydarkblue,above right=-2] at (R0') {$1$};
  \node[fill=mydarkblue,circle,inner sep=0.8] (R1') at (R1) {};
  \node[mydarkblue,above right=-2] at (R1') {$\tau$};
  \node[fill=mydarkblue,circle,inner sep=0.8] (R2') at (R2) {};
  \node[mydarkblue,above right=-2] at (R2') {$\tau^2$};
  \node[fill=mydarkblue,circle,inner sep=0.8] (R3') at (R3) {};
  \node[mydarkblue,above right=-2] at (R3') {$\tau^3$};
  \node[fill=mydarkblue,circle,inner sep=0.8] (R4') at (R4) {};
  \node[mydarkblue,above right=-2] at (R4') {$\tau^4$};
  \node[fill=mydarkblue,circle,inner sep=0.8] (R5') at (R5) {};
  \node[mydarkblue,above right=-2] at (R5') {$\tau^5$};
  
  \draw[->,line width=0.9] (-0.05,0) -- (\xmax+0.05,0) node[right] {Re};
  \draw[->,line width=0.9] (0,-0.05) -- (0,\ymax+0.05) node[left] {Im};
  \draw[dashed] (R0) -- (R1) -- (R2) -- (R3) -- (R4) -- (R5) -- (R0);
  \draw[dashed] (R0) -- (R3);
  \draw[dashed] (R1) -- (R4);
  \draw[dashed] (R2) -- (R5);

  \draw[->, blue, line width = 2.0] (O) -- node[anchor = south,align=center] {$M_1$} (R0);
  \draw[->, blue, line width = 2.0] (O) --node[anchor = south west,align=center] {$M_1$} (R2);
  \draw[->, orange, line width = 2.0] (O) --node[anchor = south,align=center] {$M_4$} (R3);
  \draw[->, blue, line width = 2.0] (O) --node[anchor = south east,align=center] {$M_1$} (R4);

  \draw[->, red, line width = 2.0] (O) --node[anchor = south,align=center] {$M_2$} (B0);
  \draw[->, red, line width = 2.0] (O) --node[anchor = south,align=center] {$M_2$} (B1);
  \draw[->, red, line width = 2.0] (O) --node[anchor = west,align=center] {$M_2$} (B2);
  \draw[->, green, line width = 2.0] (R0) --node[anchor = south,align=center] {$M_3$} (B0);
  
\end{tikzpicture}
  \caption{Definitions of monodromies $M_1$, $M_2$, $M_3$ and $M_4$}
  \label{fig:monodromies_sketch}
\end{figure}
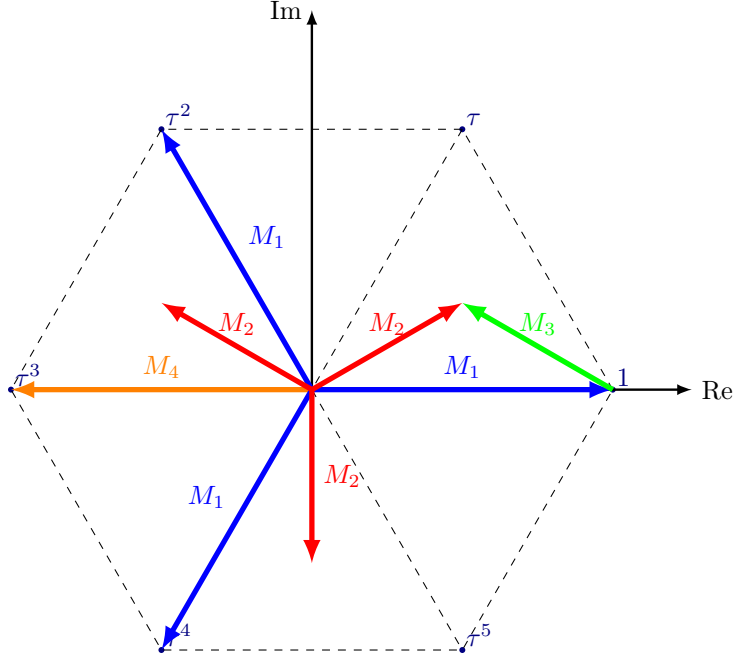

Now, consider a flat connection on $\mathbb C$ that is symmetric under translations and that also has an additional symmetry under rotations $g:z\mapsto \omega z$ that act by a pullback of the connection form. That means that monodromies $M_1$ along blue paths as shown on Fig. \ref{fig:monodromies_sketch} are equal. Monodromies $M_2$ are also equal between each other. The green path can be obtained by shifting one of the red paths by $1$. Its monodromy is then equal to $M_3 = T M_2 T^{-1}$. Because the connection is flat, we have
\beq
M_1 = M_3^{-1} M_2  = T M_2^{-1} T^{-1} M_2
\eeq

The monodromies $M_a$ and $M_b$ around $a$- and $b$-cycles of the torus also includes the transformation $T^{-1}$ needed to get back to the point of origin. Therefore,
\beq
    M_a = M_b = T^{-1} M_1 = M_2^{-1} T^{-1} M_2
\eeq

\beq
    \Tr M_a = \Tr M_b = \Tr T = -1
\eeq

Thus, flat connections with a ${\BZ}_3$ symmetry all correspond to a single point in the moduli space of flat connections $\mathcal M_{0}$.

Now, consider a connection with a larger ${\BZ}_6$ symmetry that acts according to $\eqref{eq:Z6_symmetry_action}$ (notice that this action contains the ${\BZ}_3$ group action studied above). It imposes an additional constraint on the monodromies:
\beq
M_4 = T^{-1} M_1^{-1} T = \begin{pmatrix}
    0 & i \\ -i & 0
\end{pmatrix} M_1 \begin{pmatrix}
    0 & i \\ -i & 0
\end{pmatrix}
\eeq
For a generic $SU(2)$ monodromy matrix
\beq 
M_1 = \begin{pmatrix}
    a & b \\ -b^* & a^*
\end{pmatrix}, \quad \quad |a|^2 + |b|^2 = 1
\eeq
this simplifies to
\beq
b^* = -\omega^{-1} b \implies b \propto i\omega^{-1}
\eeq
This completes the proof that in case of a ${\BZ}_6$ symmetry present in TBG, the monodromies $M_a = M_b = T^{-1}M_1$ of the flat connection at $t\to \infty$ have the form \eqref{eq:monodromy_Z6_form} and can be parameterized by $\phi \in \mathbb R / 2\pi \mathbb \BZ$. This parametrization is not gauge invariant, but its winding number around a critical point is invariant under the gauge transformations that are continuous in $\alpha$.

\end{document}